\documentclass[fleqn,usenatbib]{mnras}

\usepackage{amssymb}
\usepackage{newtxtext,newtxmath}
\usepackage[T1]{fontenc}
\usepackage{breqn}
\usepackage{multirow}
\usepackage{upgreek}

\DeclareRobustCommand{\VAN}[3]{#2}
\let\VANthebibliography\thebibliography
\def\thebibliography{\DeclareRobustCommand{\VAN}[3]{##3}\VANthebibliography}

\usepackage{graphicx}	% Including figure files
\usepackage{amsmath}	% Advanced maths commands
\usepackage{import}

\usepackage[normalem]{ulem}

\title[Structured GRB jet cooling and spreading]{Moving-mesh simulations of spreading dynamics and local electron cooling in structured gamma-ray burst afterglow jets}
\author[Kundu, van Eerten]{
Sayan Kundu$^{1}$\thanks{E-mail: sayan.astronomy@gmail.com},
Hendrik van Eerten$^{1}$
\\
$^{1}$Department of Physics, University of Bath, Bath, UK
}

\date{Accepted XXX. Received YYY; in original form ZZZ}

\pubyear{2026}

\begin{document}
\label{firstpage}
\pagerange{\pageref{firstpage}--\pageref{lastpage}}
\maketitle

\begin{abstract}
We present the results for the dynamics and emission profiles of axi-symmetric numerical simulations of structured
gamma-ray burst afterglow jets, computed using the relativistic moving-mesh hydrodynamics code GAMMA. We find that the
spreading of jets of average opening angle is moderately impacted by the initial steepness of the angular structure, although the
effect disappears once the working surface of the jet substantially exceeds its initial width, and that the travel time of a sound
wave across the front surface remains the best indicator of the onset of spreading also for structured jets. When computing the
afterglow spectrum using a local cooling approach that traces the electron population following shock-acceleration, we observe
a significant impact on the synchrotron cooling break. Similar to earlier results for top-hat jets, the cooling break is found to shift
upward in frequency by well over a factor of ten relative to approaches that assume a global cooling timescale across the jet.
The cooling break transition in the spectrum also becomes substantially smoother. For both local and global cooling, jet breaks become sharper with increasing
frequency. Local cooling is found to initially lead to a steeper slope post jet-break. The local-cooling emission is shown to originate from a narrow frequency-dependent sized region behind
the shock front, as expected, but in strong contrast to a global cooling approach.

\end{abstract}


%%%%%%%%%%%%%%%%%%%%%%%%%%%%%%%%%%%%%%%%%%%%%%%%%%

%%%%%%%%%%%%%%%%% BODY OF PAPER %%%%%%%%%%%%%%%%%%

\section{Introduction}

Gamma-ray bursts (GRBs) are among the most explosive  events in the universe. 
They are typically observed to release a vast amount of energy in a very short period of time on the order of seconds to minutes.
We still do not have a very clear understanding of how and why these events happen in the first place, although usually their origin is attributed either to core collapse of a very massive star \citep{Woosley_1993, MacFadyen_1999} or to the merging events of neutron stars \citep{Eichler_1989,Mochkovitch_1995, Abbott:2017multi-messenger}. 
Depending on the observational features, the emission from a GRB is typically divided into two broad categories: prompt and afterglow phase.
The prompt phase consists of sudden bursts, typically in gamma rays or hard X-rays, which are generally associated with internal dissipation within the outflow. Conversely, the onset of the afterglow phase is attributed to the interaction of the outermost shock front with the surrounding ambient medium and typically results in multi-wavelength emission features.
Further, on the basis of the timescale of the prompt emission phase and other characteristics such as spectral hardness, GRBs are categorised into two broad classes: short GRB (sGRB) and long GRB (lGRB), which are attributed to merging events of neutron stars and the core collapse of massive stars, respectively \citep[see e.g.][for reviews]{Zhang_2004,Kumar_2015}.

To understand the dynamics and emission characteristics of GRBs, various models have been employed.
Among them, the fireball model is one of the most celebrated due to its capability to reproduce various observational features \citep{Rees_1992}.
Comparing the model with various observational features of GRBs, the fireball is typically assumed to have a jet-like structure instead of a sphere-like morphology.
Such conical jet flow, with time, will start to spread in the lateral direction and eventually achieve a spherical structure. 
The spreading behaviour of the jet is also known to have an impact on the afterglow emission \citep{Rhoads_1997, Rhoads_1999, Sari_1999}.

Being cosmologically distant sources whose emission is strongly beamed along the direction of their relativistic motion, GRBs are mostly observed on-axis, with one of their jets pointed straight towards the observer. At this orientation, the impact of the lateral distribution of energy across the jet on observations is only minor, provided that the jet's uniform core holds the majority of its total energy. Consequently, models that extend the simple picture of angle-independent outflow energy (a "top-hat" jet profile) to one where the jets contain additional structure have long existed in the literature (e.g. \citealt{Rossi_2002, Kumar_2003, Lamb_2005, Lazzati2005}, see also \citealt{vanEerten_2018} for a review).

The need for incorporating structure in GRB jet models was reinforced dramatically by the joint discovery of GRB 170817A and GW170817, the first (and so far only) GRB accompanied by a direct detection of gravitational waves produced by the merging of two neutron stars. Over the course of the evolution of the afterglow, the evidence for a structured jet seen off-axis became increasingly compelling (see e.g. \citealt{Troja_2017, Margutti2017ApJ, Hallinan_2017, Mooley_2018, LambKobayashi2018MNRAS, gill_2018, gill_2019numerical,Ghirlanda2019, Troja_2019, Troja_2020, Ryan_2020} and many others) and there is currently a consensus that the rising slope of the light curve in particular reflects how the emission from emission from different annuli of the jet profile enters into our line of sight (e.g. \citealt{Ryan_2020, TakahashiIoka2020MNRAS, beniamini_2020afterglow}).

Since the first papers \citep{Rhoads_1997, Rhoads_1999, Sari_1999}, the spreading dynamics of jets have been studied numerically \citep{granot_2001_grba,Kumar_2003, Meliani_2007, ZhangMacFadyen_2009, vanEerten_2010a, 
Wygoda2011, vanEerten_2012, eerten_2012_boxfit, Lyutikov2012, eerten_2013_boosted, DeColle_2012, DeColle2012ApJdynamics} and modelled (semi-)analytically \citep{Gruzinov2007, GranotPiran_2012, Keshet2015, DuffellLaskar2018ApJ, Ryan_2020}. More recent work by \cite{Govreen-Segal_2024}, using the same relativistic hydrodynamics code as used throughout this paper, has extended this study to narrow structured jets. Their work has demonstrated that sufficiently narrow structured jets are able to achieve an intermediate stage of ultra-relativistic spreading where previous demonstrations were limited to top-hat jets (see  \citealt{vanEerten2013eConf} for discussion).

An important aspect of GRB afterglow modelling is the treatment of electron cooling. 
Early simplified analytical estimates by \cite{Sari_1998} approximated the material downstream of the blast wave shock front as a single zone, where a balance is achieved between synchrotron cooling,  injection of non-thermal particles and adiabatic expansion. A \textit{global} timescale for the cooling is assumed, taken equal to the adiabatic expansion timescale (which in turn can be conflated with the time since the explosion triggering the blast wave).

A more realistic, \textit{local} cooling prescription can be implemented using a full description of the blast wave radial profile, as was done by \cite{Granot_1999, granot_2002}. 
Later work  \citep{Guidorzi_2014,vanEerten_2010a} confirmed that there is a significant offset between the cooling frequencies computed from local and global cooling. Furthermore, the smoothness of the cooling break established by \cite{GranotSari_2002} for a local cooling approach was confirmed by \cite{uhm2014}, who emphasized the role of changing magnetic field and other physical conditions of the cooling plasma dropping further downstream of the shock front. 

In a local approach to electron cooling, the region behind the shock front is hot enough to radiatiate substantially at high (e.g. X-ray) frequencies becomes very small at the high end of the electron spectrum, due to the steep drop in electron energy with time since shock-acceleration. Local electron cooling therefore poses a numerical challenge similar to that posed by the extreme thinness of the blast wave at high Lorentz factors. For this reason, afterglow emission computed from relativistic hydrodynamics jet simulations has so far predominantly been done through a hybrid approach \citep{ZhangMacFadyen_2009, vanEerten_2010a, eerten_2012_boxfit, DeColle2012ApJdynamics}. Based on a global timescale (again, time since explosion), the emission is then still computed locally for each region of the flow. With the more recent development of moving-mesh hydrodynamics codes better able to resolve the thin hot region behind the shock front \citep{Duffell2013ApJ, Ayache_2020, Ayache_2022}, the first simulation have been published that trace electron cooling locally \citep{Ayache_2022} in a multi-dimensional (axi-symmetric) simulation. However, \cite{Ayache_2022} assumed a top-hat jet profile and the impact of different cooling prescriptions on structured GRB jets remains so far unexplored. 

In this work, we investigate how local and global cooling shape the afterglow emission of structured jets using moving-mesh axi-symmetric relativistic hydrodynamic simulations and explore the spreading dynamics of structured jets where the angular structure is modelled either with a Gaussian or a power-law distributions of energy with outflow angle. The paper is structured as follows. We introduce our numerical approach and our choice of initial conditions for the jet dynamics in section~\ref{sec:numerics}. In section \ref{sec:emission} we introduce our approach to computing radiation and modelling electron cooling locally and globally. The dynamical results from the simulations are presented in section~\ref{sec:results_dyn}, while the emission results are described in section~\ref{sec:rad}.
We summarise and discuss the implication of both in section~\ref{sec:discuss}, along with the limitations of this work and possible future extensions.
We conclude the work in section~\ref{sec:conclusion}.

\section{Numerical method for jet dynamics}{\label{sec:numerics}}

\begin{figure*}
    \centering
    \includegraphics[width=\textwidth]{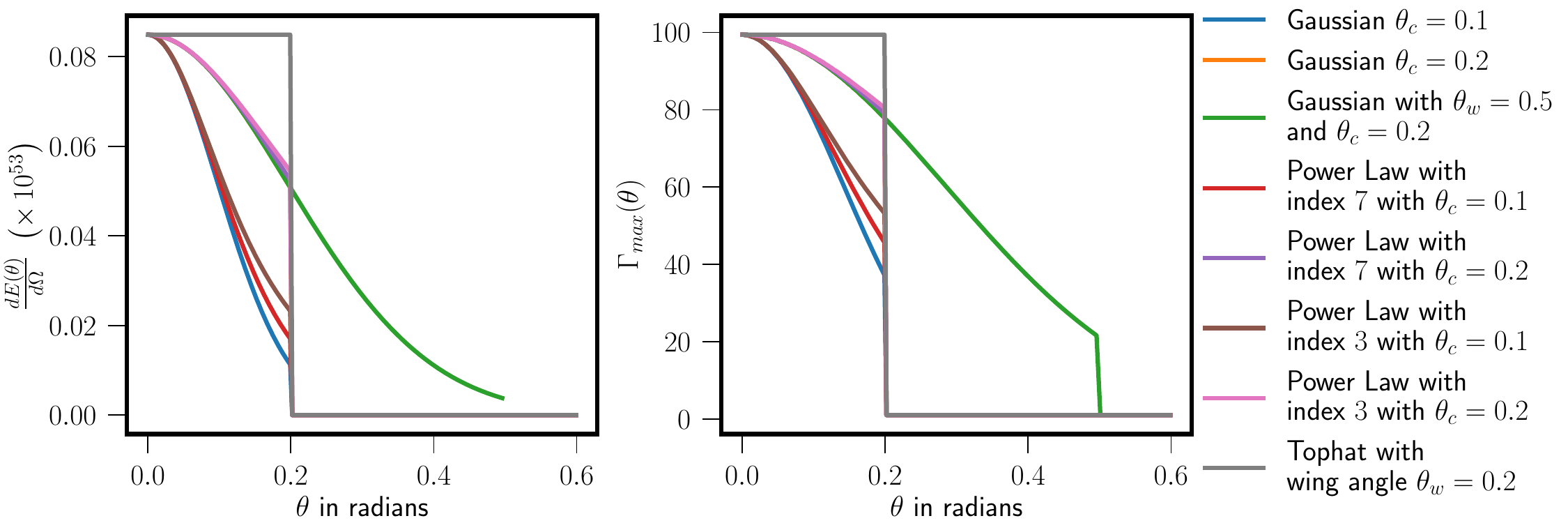}
    \caption{Initial angular distribution of Energy (\emph{left}) and maximum fluid Lorentz factor (\emph{right}) for different jet structures with different core angles, $\theta_{\rm c}$. 
    The wing angle, $\theta_{\rm w}$, for all the simulations is kept constant with a value of $0.2$ radian.}
    \label{fig:initial}
\end{figure*}

The appearance of a GRB afterglow is dictated by the relativistic dynamics of the forward shock and its interaction with the circumburst medium.
Therefore, a proper numerical study of GRB afterglow from structured jet demands resolving this relativistic forward shock with sufficient precision and tracking its evolution until the dynamics become non-relativistic. 
For this purpose, we employ the special relativistic hydrodynamic (SRHD) code \rm{GAMMA}\footnote{Publicly available at \url{https://github.com/eliotayache/GAMMA}.} \citep{Ayache_2020, Ayache_2022}, which has been developed utilising a moving mesh concept following an approach first applied in the context of relativistic jets by \cite{Duffell2013ApJ}.
In particular, we have carried out axi-symmetric jet simulations in a polar geometry $\{r,\theta\}$ by solving the following set of conservation equations,
\begin{equation}\label{eq:continuity}
    \frac{\partial \left(\rho\gamma\right)}{\partial t} + \frac{1}{r^{2}}\frac{\partial\left(r^{2}\rho \gamma v_{\rm r}\right)}{\partial r}+\frac{1}{r\sin\theta}\frac{\partial \left(\rho \gamma v_{\uptheta}\sin\theta \right)}{\partial \theta}=0,
\end{equation}
\begin{dmath}\label{eq:momentumr}
    \frac{\partial \left(\rho h\gamma^{2}v_{\rm r}\right)}{\partial t} + \frac{1}{r^{2}}\frac{\partial\left(r^{2}\rho h\gamma^{2}v_{\rm r}^{2}\right)}{\partial r}+
    \frac{1}{r\sin\theta}\frac{\partial \left(\rho h\gamma^{2}v_{\rm r} v_{\uptheta}\sin\theta \right)}{\partial \theta}+\frac{\partial p}{\partial r}=\frac{\rho h\gamma^{2}v_{\uptheta}^{2}}{r},
\end{dmath}
\begin{dmath}\label{eq:momentumt}
    \frac{\partial \left(\rho h\gamma^{2}v_{\uptheta}\right)}{\partial t} + \frac{1}{r^{2}}\frac{\partial\left(r^{2}\rho h\gamma^{2}v_{\rm r}v_{\uptheta}\right)}{\partial r}+
    \frac{1}{r\sin\theta}\frac{\partial \left(\rho h\gamma^{2}v_{\uptheta}^{2}\sin\theta \right)}{\partial \theta}+\frac{1}{r}\frac{\partial p}{\partial \theta}=-\frac{\rho h\gamma^{2}v_{\rm r}v_{\uptheta}}{r},
\end{dmath}
\begin{dmath}\label{eq:energy}
    \frac{\partial }{\partial t} \left(\rho h\gamma^{2}-p-\rho \gamma\right) +     \frac{1}{r^{2}}\frac{\partial\left(r^{2}\rho h\gamma^{2}v_{\rm r}-\rho \gamma \right)}{\partial r}+ \frac{1}{r\sin\theta}\frac{\partial \left[\left(\rho h\gamma^{2}v_{\uptheta}-\rho\gamma\right)\sin\theta \right]}{\partial \theta}=0,
\end{dmath}
where $\rho$, $\gamma$, $p$, $v_{\rm r}$ and $v_{\uptheta}$ represent the density, Lorentz factor, pressure, radial velocity, and lateral velocity, respectively. $h$ is the specific enthalpy including rest mass energy and is defined as $\rho h=\rho c^2 + p + e$, with $e$ the internal energy density (excluding rest mass term) and $c$ the speed of light.
In addition to the above equations, we consider an ultra-relativistic equation of state (EOS), suited for an ideal relativistic monoatomic gas to close the SRHD equations, which relates pressure and internal energy as follows:
\begin{equation}\label{eq:eos}
    p = e \left( \hat{\gamma}-1\right) \,\,\text{with} \,\, \hat{\gamma} =\frac{4}{3}.
\end{equation}
Numerically, Eqs.~\ref{eq:continuity}-\ref{eq:energy} are solved using the Harten, Lax, van Leer contact (HLLC) Riemann solver \citep{mignone_2005_hydroHLLC} adopting a piecewise linear reconstruction with minmod flux limiter.
Further, a $3^{rd}$ order Runge-Kutta time-stepping algorithm is employed for the temporal evolution of the variables.

\textsc{GAMMA} uses a mesh refinement technique in addition to the moving-mesh architecture that helps to follow the dynamics of the shock for a longer time.
The mesh refinement is dictated by a ``re-gridding score'' $S_{\rm g}=a\gamma^{3/2}$ with $a=dr/r_{\rm max}d\theta$ and $\gamma$ the local fluid Lorentz factor. This definition of $a$ prevents the time step from being governed by cells located at smaller radii at the late evolution times.
The variation of $S_{\rm g}$ is further constrained to lie in the interval $[0.1,3]$.

\subsection{Initial conditions setting jet dynamics}
We have carried out axi-symmetric simulations of structured GRB jets with a lateral resolution $N_{\uptheta} = 300$, covering a domain extending from $0$ to $\pi/2$ radians using tracks of varying width $\Delta \theta$. To this end, we set the position of the interface between the $j^{th}$ and $(j-1)^{th}$ track as follows,
\begin{equation}
    \theta_{\rm j-1/2}=\frac{\pi}{2}\left(0.3\left(\frac{j}{N_{\uptheta}}\right)+0.7\left(\frac{j}{N_{\uptheta}}\right)^{3}\right)
\end{equation}
The radial extent of the simulation box depends on the initial position of the forward shock, which we calculate by specifying the energy contained in each track $E(\theta)$ and the desired fluid Lorentz factor at the immediate downstream according to our initial jet radial profile.

This profile is determined using the self-similar solution from \cite{blandford_1976} applicable to the relativistic limit of the strong point explosion problem (hereafter abbreviated as the ``BM'' solution). At early times, the jet flow is expected to remain approximately radial due to a lack of causal contact across the jet, rendering this self-similar solution applicable even to simulations in non-spherical symmetry. In this study, we assume a homogeneous circumburst medium of density $\rho_{\rm ext}$. Along a given radial line, the BM solution relates $\rho_{\rm ext}$, shock Lorentz factor $\Gamma_{\rm s} \left(\theta \right)$, isotropic-equivalent (for that angle) explosion energy $E_{\rm iso} \left( \theta \right)$ and simulation lab frame time $t$ according to
\begin{equation}\label{eq:energy_BM}
    t = \left(\frac{17 E_{\rm iso}}{8\pi\rho_{\rm ext} \Gamma_{\rm s}^{2}c^{5}}\right)^{\frac{1}{3}}.
\end{equation}
We use this relation for a given $E_{\rm iso} \left( \theta = 0 \right)$ and $\Gamma_{\rm s} \left( \theta = 0 \right)$ to determine the initial time $t_{\rm i}$ of the simulation and then determine $\Gamma_{\rm s} \left( \theta \right)$ from a fixed time $t = t_{\rm i}$ and the desired jet profile $E_{\rm iso} \left( \theta \right)$. Because the BM solution applies to a point-release of pure energy, our simulations do not include a deceleration phase of the blast wave (or, equivalently, a separate constraint on initial Lorentz factor). Instead, our starting values for $\Gamma_{\rm s}$ merely represent the starting time of the simulation.

The radius $R_{\rm s}$ of the forward shock is then calculated from the BM solution via the following equation:
\begin{equation}\label{eq:shock_rad}
    R_{\rm s}(\theta) = \left(1-\frac{1}{1+8\Gamma_{\rm s}^{2}}\right) c t_{\rm i}.
\end{equation}
Consequently, we set the radial extent of the track using
\begin{equation}
r\in\left(R_{\rm s}-50R_{\rm s}/\Gamma_{\rm s}^{2},R_{\rm s}+50R_{\rm s}/\Gamma_{\rm s}^{2}\right),
\end{equation}
and discretise it with $N_{\rm r}=9900$ cells.
The fluid profile downstream of the shock is discretised with a lower cell width compared to the upstream with the aim to properly resolve the blast wave width $\Delta R_{\rm s}\simeq R_{\rm s}/ 6\Gamma_{\rm s}^{2}$.

According to the BM solution, we set the fluid parameters (pressure $p$, bulk Lorentz factor $\gamma$, mass density $\rho$ for the discretized grid in the downstream region according to
\begin{equation}\label{eq:BM}
    \begin{aligned}
        \chi &= \left[1+8\Gamma_{\rm s}^{2}\right]\left(1-\frac{r}{ct}\right), \\
        \rho \gamma &= 2 \rho_{\rm ext}\Gamma_{\rm s}^{2}\chi^{-7/4}, \\
        \gamma &= \left( 2 \Gamma_{\rm s}^2 \chi^{-1} + 1 \right)^{1/2}, \\
        p &= \frac{2}{3}\rho_{\rm ext}c^{2}\Gamma_{\rm s}^{2}\chi^{-17/12},
    \end{aligned}
\end{equation}
where $\chi$ the self-similar coordinate defined in the region $r < R_{\rm s} \left(\theta \right)$. The expression for $\gamma$ includes a second term to ensure that $\gamma \ge 1$ everywhere on the grid (the BM solution is a first-order approximate solution in $1/\gamma^2$; this limitation needs to be taken into account where $\gamma \sim 1$ even if these regions do not contribute to the overall flow dynamics). 

The fluid parameters in the upstream region are kept at fixed $\rho_{\rm ext}=n_{\rm ext}m_{\rm p}$ and $p_{\rm ext}=\eta\rho_{\rm ext}c^{2}$ with $n_{\rm ext}$, $\eta$ and $m_{\rm p}$ being the number density of the external medium, the relativistic temperature of the external medium and the mass of the proton, respectively.

\begin{table}
\centering
\caption{
Jet structure models and angular core sizes used in the simulations. The tip energy is kept at $E_{\rm 0} = 10^{53}$\,erg. Further, note that there is no concept of the core angle in the tophat jet.
}
\begin{tabular}{lllll}
\hline
label & Jet structure & Expression & $\theta_{\rm c}$ & $\theta_{\rm w}$\\
 & & & (rad) & (rad) \\
\hline
G-12 & Gaussian  & $E_0 e^{-\theta^2 / 2\theta_{\rm c}^2}$ & 0.1 & 0.2 \\
G-22 &                           &                                   & 0.2 & 0.2 \\
G-25 &                          &                                   & 0.2 & 0.5 \\
PL3-12 & Power law (index 3) & $E_{\rm 0} \left(1 + \frac{\theta^2}{3\theta_{\rm c}^2}\right)^{-\frac{3}{2}}$ & 0.1 & 0.2 \\
PL3-22 &   &                                                              & 0.2 & 0.2 \\
PL3-055 &   &   & 0.05 & 0.5 \\

PL7-12 & Power law (index 7) & $E_{\rm 0} \left(1 + \frac{\theta^2}{7\theta_{\rm c}^2} \right)^{-\frac{7}{2}}$ & 0.1 & 0.2 \\
PL7-22 &  &                                                              & 0.2 & 0.2 \\
TH-2 & Top hat & $E_{\rm 0}$ & -- & 0.2 \\
\hline
\end{tabular}
\label{tab:1}
\end{table}

\begin{table}
    \centering
    \caption{Simulation parameters for dynamics and radiation used in this study.}
    \label{tab:parameters}
    \begin{tabular}{ll}
        \hline
        Parameter & Value \\
        \hline
        Circumburst density $n_{\rm ext}$ & $1.0$ cm$^{-3}$ \\
        Circumburst temperature $\eta \equiv p / ( \rho c^2)$ & $10^{-5}$ \\
        Initial time $t_{\rm i}$ & $4.39 \times 10^6$\,s \\
        Fluid Lorentz factor behind the shock & $100$ \\
        Fractional energy in electrons $\epsilon_{\rm e}$ & $0.1$ \\
        Fractional energy in magnetic field $\epsilon_{\rm B}$ & $0.01$ \\
        Initial energy slope injected electrons $p$ & 2.23 \\ 
        \hline
    \end{tabular}
\label{tab:2}
\end{table}

In order to explore the effect of structure on jet dynamics and emission, we set up a series of jet structure profiles as specified in Table \ref{tab:1}. 
For the power-law jet we take
\begin{equation}\label{eq:general_prof}
E(\theta) = E_{\rm 0}\left(1+\frac{\theta^{2}}{n\theta_{\rm c}^{2}}\right)^{-n/2},
\end{equation}
with pre-factor $E_0$, characteristic width $\theta_c$ and power-law exponent $n$.
The jet is truncated at an angle $\theta_{\rm w}$. The Gaussian and top-hat jet models ($E(\theta) \propto \exp(-\theta^2 / 2 \theta_{\rm c}^2)$ and $E(\theta) \propto \theta^0$, respectively) represent asymptotic limits of the expression
for high and low values of parameter $n$, albeit that the ratio $\theta_{\rm w} / \theta_{\rm c}$ controls the extent to which the structure of the energy drop-off dominates the jet profile in the wings.
This functional form for the energy profile has also been used in the afterglow modelling software package \texttt{Afterglowpy}\footnote{Publicly available at \url{https://github.com/geoffryan/afterglowpy}.} \citep{Ryan_2020, Ryan_2024ApJ}.
In Table~\ref{tab:2} we provide all the values of the parameters chosen to specify initial fluid conditions and used in the radiation calculation (discussed in Section \ref{sec:emission}).  

Fig.~\ref{fig:initial} shows the resulting initial energy profile for the different jets. This profile is computed from the simulation grid using the equation
\begin{equation}
\frac{dE \left( \theta \right)}{d\Omega} = \frac{E_{\rm iso} \left( \theta \right)}{4 \pi} = \int r^{2}\tau dr,
\end{equation}
where $\tau \equiv \gamma^2 \rho h  - p - \gamma \rho$ is the relativistic energy density of the jet (not counting rest mass).

\section{Numerical method for synchrotron emission}
\label{sec:emission}

To compute the radiation from our dynamical simulations, we have used the linear radiative transfer code by \cite{Eerten_2009, eerten_2012_boxfit}. This solves the linear radiative transfer problem of synchrotron emission, computing
\begin{equation}
\frac{d I_{\upnu}}{d s} = j_{\upnu} - \alpha_{\nu} I_{\upnu},
\end{equation}
simultaneously for a large number of rays passing through the evolving jet. Here $j_{\upnu}$ is the synchrotron emission coefficient for a power-law accelerated electron population $n_{\rm e}(\gamma_{\rm e})$ with slope $-p$ (not to be confused with fluid pressure $p$) in electron Lorentz factor $\gamma_{\rm e}$ (in the frame co-moving with the fluid), such that electrons are injected with $n_{\rm e} (\gamma_{\rm e}) \propto \gamma_{\rm e}^{-p}$ at $\gamma_{\rm e} > \gamma_{\rm m}$. The synchrotron self-absorption coefficient $\alpha_{\upnu}$ does not feature in the results presented in this work and will not be discussed further. 

The emission and absorption coefficients are derived from the local state of the fluid, in particular the co-moving number density $n$, fluid bulk Lorentz factor $\gamma$ and internal energy density $e$. These are combined with the assumption required for the production of synchrotron radiation that a fraction $\epsilon_{\rm B}$ of the internal energy density $e$ resides in shock-generated magentic fields such that $B^2 / (8 \pi) = \epsilon_{\rm B} e$. In the absence of electron cooling, the emission coefficient is given by
\begin{equation}
j_{\upnu} = j_{\upnu,\rm P} \times \mathcal{F} \left( \nu \right),
\end{equation}
where the pre-factor $j_{\upnu,\rm P}$ is given by
\begin{equation}
j_{\upnu, \rm P} = 9.6323 \times \frac{\sqrt{3} q_e^3}{8 \pi m_{\rm e} c^2} \frac{p-1}{3p-1} \xi_{\rm N} n B,
\end{equation}
where $m_{\rm e}$ the electron mass and $q_{\rm e}$ the electron charge, while the term $\mathcal{F} \left( \nu \right)$ encodes the power-law nature of the synchrotron spectrum as dictated at frequency $\nu$ by the lower cut-off electron Lorentz factor $\gamma_{\rm m}$ of the accelerated distribution and a second turning point associated with electron cooling. We provide our implementations of $\mathcal{F}$ and discuss the impact of electron cooling separately in the following sections for our different approaches to electron cooling. 

For a given observer time $t_{\rm obs}$ at cosmological redshift $z$, values for $j_{\upnu}$ during integration of rays directed to the observer are provided by the output of the \textsc{gamma} simulations (`data dumps', or `snapshots') at a range of simulation emission times $t$ sampled at discrete points that satisfy 
\begin{equation}\label{eq:EDS}
    \frac{t_{\rm obs}}{1+z} = t - \frac{r\mu(\theta, \phi)}{c}.
\end{equation}
Here, $r$ denotes the radial position of the fluid element being probed and $\mu$ the cosine of the angle between the fluid element position and the line-of-sight to the observer, centred at the origin of the explosion\footnote{For given fixed observer time $t_{\rm obs}$ and fixed emission time $t$, equation \ref{eq:EDS} defines an \emph{equidistant surface} containing all points from which emission released at $t$ will arrive at $t_{\rm obs}$. Had the jet been approximated as an infinitesimally thin shell, all non-zero contributing coordinate pairs $( t, r, \mu)$ would have jointly traced out an \emph{equal-arrival time surface} akin to that discussed e.g. in \cite{sari_1998_grb_shape}, with $t$ still varying across this surface. For an on-axis observer, $\mu = \cos \theta$.}.

\subsection{Global Cooling}

The synchrotron spectrum can be related to the local fluid state at varying levels of approximation to the critical electron Lorentz factors $\gamma_{\rm m}$ from particle injection and a second factor $\gamma_{\rm c}$ related to electron cooling. Traditionally, afterglow computations based on multi-dimensional SRHD simulations (e.g. \citealt{Zhang_2009, vanEerten_2010a, DeColle_2012}) have followed the approach to $\gamma_{\rm m}$ from early works on afterglows \citep{Meszaros_1994ApJ, Wijers_1997MNRAS,  Sari_1998}, where the lower cut-off follows from the assumption that a fraction $\epsilon_{\rm e}$ of $e$ resides in the non-thermal electron such that
\begin{equation}
\gamma_{\rm m} \equiv \frac{\bar{\epsilon}_{\rm e} e}{\xi_{\rm N} n m_{\rm e} c^2},
\label{eq:gamma_m}
\end{equation}
where $\bar{\epsilon}_{\rm e} \equiv \left( p - 2\right) \epsilon_{\rm e} / \left( p - 1\right)$ absorbs some factors of $p$ to facilitate generalizing the approach to $p < 2$ if needed and $\xi_{\rm N}$ the fraction of total electrons accelerated into a power-law distribution (we set $\xi_{\rm N} \equiv 1$ for simplicity; our results can be scaled to other values by making use of the model degeneracy between $\xi_{\rm N}$, $E_{\rm iso}$, $\rho_{\rm ext}$, $\epsilon_{\rm e}$, $\epsilon_{\rm B}$ pointed out by \citealt{EichlerWaxman2005}). Eq. \ref{eq:gamma_m} only remains true in the downstream medium in the absence of strong cooling affecting electrons at these Lorentz factors and when assuming that an ultra-relativistic EOS remains applicable to at least the electron population \citep{vanEerten_2010}.

The ``traditional'' treatment of electron cooling in simulation-derived light curve computations is to take an approach again going back to \cite{Sari_1998} and similar works, where a single global timescale is assumed to set the balance between adiabatic expansion and synchrotron cooling in a steady-state approximation to the evolution of the electron population. If this timescale is taken to be the lab frame time $t$ since the explosion (itself a reasonably approximation to the adiabatic expansion timescale, given that blast wave radius $R \sim c t$) and local magnetic field $B$ is used for the synchrotron losses, a hybrid expression for $\gamma_{\rm c}$ results of the form
\begin{equation}\label{eq:gamc}
    \gamma_{\rm c}=\frac{6\pi m_{\rm e}c \gamma}{\sigma_{\rm T}B^{2}t},
\end{equation}
where $\sigma_{\rm T}$ the Thomson cross section.

To each electron Lorentz factor $\gamma_{\rm e}$, including $\gamma_{\rm m}$ and $\gamma_{\rm c}$, a synchrotron critical frequency $\nu'_{\rm cr}$ (primed, to emphasize it is expressed co-moving with the radiating fluid rather than in the observer's frame) can be associated according to standard synchrotron theory via
\begin{equation}
\nu'_{\rm cr} \equiv \frac{3 q_{\rm e}}{4 \pi m_{\rm e} c} \gamma_{\rm e}^2 B.
\end{equation} 
Each fluid cell is now assumed to produce a steady-state synchrotron emission spectrum. If $\gamma_{\rm c} > \gamma_{\rm m}$, the local steady-state electron population obeys $n_{\rm e}(\gamma) \propto \gamma_{\rm m}^{-p}$ for $\gamma_{\rm m} < \gamma_{\rm e} < \gamma_{\rm c}$ and $n_{\rm e}(\gamma_{\rm e}) \propto \gamma_{\rm m}^{-p-1}$ for $\gamma_{\rm m}$, $\gamma_{\rm c} < \gamma_{\rm e}$, leading to a local synchrotron spectrum of the form
\begin{equation}
\mathcal{F} = \left\{ \begin{array}{ll} \left( \frac{\nu}{\nu^*_{\rm m}} \right)^{\frac{1}{3}} & \textrm{ for } \nu < \nu^*_{\rm m} < \nu^*_{\rm c}, \\ \left( \frac{\nu}{\nu^*_{\rm m}} \right)^{\frac{1-p}{2}} & \textrm{ for } \nu^*_{\rm m} < \nu < \nu^*_{\rm c} , \\ \left( \frac{\nu^*_{\rm c}}{\nu^*_{\rm m}} \right)^{\frac{1-p}{2}} \left( \frac{\nu}{\nu^*_{\rm c}} \right)^{-\frac{p}{2}} & \textrm{ for } \nu^*_{\rm m} < \nu^*_{\rm c} < \nu. \end{array} \right.
\end{equation}
Here $\nu^*_{\rm m}$ is the critical frequency corresponding to $\gamma_{\rm e} = \gamma_{\rm m}$, expressed in the observer frame, and likewise for $\nu^*_{\rm c}$ and $\gamma_{\rm c}$ (the asterisks are to distinguish between local critical frequencies and the volume-integrated spectral features $\nu_{\rm m}$ and $\nu_{\rm c}$ seen by the observer). If $\gamma_{\rm c} < \gamma_{\rm m}$, we have instead $n_{\rm e}(\gamma_{\rm e}) \propto \gamma_{\rm e}^{-2}$ for $\gamma_{\rm c} < \gamma_{\rm e} < \gamma_{\rm m}$ and again $n_{\rm e}(\gamma_{\rm e}) \propto \gamma_{\rm m}^{-p-1}$ for $\gamma_{\rm m}$, $\gamma_{\rm c} < \gamma_{\rm e}$, now leading to a local synchrotron spectrum of the form
\begin{equation}
\mathcal{F} = \left\{ \begin{array}{ll} \left( \frac{\nu}{\nu^*_{\rm c}} \right)^{\frac{1}{3}} & \textrm{ for } \nu < \nu^*_{\rm c} < \nu^*_{\rm m}, \\ \left( \frac{\nu}{\nu^*_{\rm c}} \right)^{-\frac{1}{2}} & \textrm{ for } \nu^*_{\rm c} < \nu < \nu^*_{\rm m} , \\ \left( \frac{\nu^*_{\rm m}}{\nu^*_{\rm c}} \right)^{-\frac{1}{2}} \left( \frac{\nu}{\nu^*_{\rm m}} \right)^{-\frac{p}{2}} & \textrm{ for } \nu^*_{\rm c} < \nu^*_{\rm m} < \nu. \end{array} \right.
\end{equation}
This global cooling approach is shared by simulation-derived light curve generating tool \textsc{boxfit} \citep{vanEerten_2012} and single-zone analysis software such as \texttt{Afterglowpy} \citep{Ryan_2020, Ryan_2024ApJ}.

The cooling break from the combined emission of the entire fluid seen by the observer will be an averaged result shaped by the integrated spectra featuring individual $\nu_{\rm c}^*$ values.

\subsection{Local Cooling}

In a local cooling scenario, we instead trace explicitly the bracketing Lorentz factors $\gamma_{\rm m}$ and $\gamma_{\rm max}$ of an injected power-law population of electrons. For this purpose, we employ the inbuilt shock detection algorithm \citep{rezzolla_2003, zanotti_2010, Ayache_2022}, and the non-thermal particle population tracking of \textsc{GAMMA}, which work as follows. Once a cell undergoes a shock, \textsc{GAMMA} detects this and assigns a $\gamma_{\rm m}$ and $\gamma_{\rm max}$ value to that cell where $\gamma_{\rm m}$ is computed according to Eq. \ref{eq:gamma_m} and upper limit $\gamma_{\rm max}$ is set to a numerically large value of $10^8$. Physically, the upper cut-off $\gamma_{\rm max}$ should reflect the upper limit achievable by shock-acceleration (where acceleration and synchrotron loss timescales are in balance), or just taken to be infinity for simplicity (as done by e.g. \citealt{GranotSari_2002}). Once set, $\gamma_{\rm max}$ is physically expected to rapidly drop to a lower value due to the impact of synchrotron cooling, which is proportional in strength to $\gamma_{\rm e}^2$. In practice \citep{Ayache_2022}, $10^8$ is found to be a good compromise between a value low enough to be tractable numerically and high enough not to arbitrarily affect emission computations at our frequency range of interest (i.e. up to X-rays).

As the simulation proceeds, the $\gamma_{\rm m}$ and $\gamma_{\rm max}$ (or any $\gamma_{\rm e}$, for that matter) of a previously shocked cell evolve according to the following equation due to synchrotron cooling and adiabatic expansion/contraction:
\begin{align}\label{eq:loss}
 \frac{d \gamma_{\rm e}}{d t^\prime} &=
        - \frac{\sigma_{\rm T} B^2}{6 \pi m_{\rm e} c } \gamma_{\rm e}^2
        + \frac{\gamma_{\rm e}}{3 \rho}\frac{d \rho}{d t^\prime},
\end{align}
where $t'$ is the time in the frame co-moving with the fluid.  Eq.~\ref{eq:loss} can be written as a passive scalar equation in the following way \citep{Downes_2002, vanEerten_2010},
\begin{dmath}
    \label{eq:advect_cool}
    \frac{\partial}{\partial t} \left( \frac{\Gamma \rho^{4/3}}{\gamma_{\rm e}} \right) + \frac{1}{r^{2}}\frac{\partial}{\partial r} \left( \frac{\Gamma \rho^{4/3}}{\gamma_{\rm e}} v_{\rm r} \right)+\frac{1}{r\sin\theta}\frac{\partial}{\partial \theta}\left(\frac{\Gamma \rho^{4/3}}{\gamma_{\rm e}}\sin\theta v_{\uptheta}\right)
      = \frac{\sigma_{\rm T}}{6 \pi m_{\rm e} c} \rho^{4/3} B^{2},
  \end{dmath}
and is solved for $\gamma_{\rm m}$ and $\gamma_{\rm max}$ in addition to Eqs.~(\ref{eq:continuity})-(\ref{eq:energy}).

The local emission is now no longer computed assuming a steady-state solution to the electron distribution. Instead, we have (with $\nu^*_{\rm max}$ the critical frequency in the observer frame corresponding to $\gamma_{\rm max}$, analogous to $\nu^*_{\rm m}$ and $\nu^*_{\rm c}$ before):
\begin{equation}
\mathcal{F} = \left\{ \begin{array}{ll} \left( \frac{\nu}{\nu^*_{\rm m}} \right)^{\frac{1}{3}} & \textrm{ for } \nu < \nu^*_{\rm m} \textrm{, } \nu^*_{\rm max}, \\ \left( \frac{\nu}{\nu^*_{\rm m}} \right)^{\frac{1-p}{2}} & \textrm{ for } \nu^*_{\rm m} < \nu < \nu^*_{\rm max} , \\ 0 & \textrm{ for } \nu^*_{\rm max} < \nu. \end{array} \right.
\end{equation}
In this approach, the eventually observed cooling break $\nu_{\rm c}$ is again an emergent feature that follows once all simultaneously arriving local contributions are integrated. Different here is that the post-cooling slopes $-1/2$, $-p/2$ are \emph{also} fully emergent features that do not appear locally. A fully analytical example of this approach, based on the Blandford-McKee self-similar solution, can be found in \cite{GranotSari_2002}. 

One striking contrast between the local and global cooling approaches as defined above, is that in local cooling the evolution of the non-thermal particle distribution is determined by the time elapsed since the last shock encounter of that zone, whereas in global cooling the non-thermal distribution throughout the entire volume is computed based on the age of the system. This distinction makes local cooling more suitable for systems comprised of multiple shocks, as it can reveal variations in emission properties arising from complex shock networks in various astrophysical systems \citep[see][for example]{kundu_2022,giri_2022}. 
In this work, we focus on understanding the difference that these two cooling prescriptions have on the emission properties of structured GRB jets with a single forward shock. For a top-hat jet, such an analysis has already revealed a significant shift in the cooling break frequency computed from the spectral energy distribution (SED) based on these two scenarios \citep{Ayache_2022}.

As mentioned previously, our choices for radiation parameters $\epsilon_{\rm B}$, $\epsilon_{\rm e}$ and $p$ are provided in Table \ref{tab:2}.

\section{Results for the dynamics of Structured jets}\label{sec:results_dyn}

Here we present the results from the RHD simulations of different jet profiles.
This section focuses on understanding the evolution of different dynamical parameters for various jet structures, while in the next section, we present the emission signatures that we obtain from these dynamical runs.

\subsection{Code verification}

\begin{figure}
    \centering
    \includegraphics[width=\columnwidth]{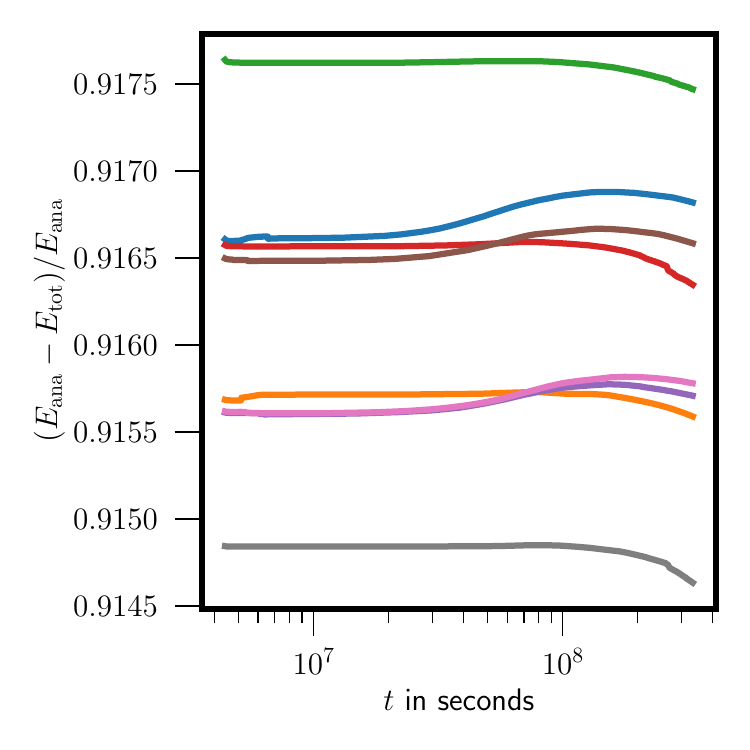}
    \caption{A consistency check on the total jet energy evolution for the different simulated jets, showing the fractional deviation of simulation grid energy $E_{\rm tot}$ from the theoretically expected value $E_{\rm ana}$.
    The color coding is as indicated in the legend of Fig. \ref{fig:initial}, i.e. top to bottom: G-25 (green), TH-2 (grey), PL3-22 (pink), PL7-22 (purple), G-22 (orange), PL3-12 (brown), PL7-12 (red), G-12 (blue).}
    \label{fig:energy_cons}
\end{figure}

We first test whether our simulations correctly preserve the energy of the blast wave during their run and that e.g. not too much energy gets lost through the movement of the boundaries. Fig. \ref{fig:energy_cons} confirms how the total energy content of the jets can be seen to remain constant with time. The total energy of each jet captured by the simulations at their current resolution, lies within 9\% of its theoretical value, changing by less than 0.05\% during a run.

\begin{figure}
%   \centering
    \includegraphics[width=\columnwidth]{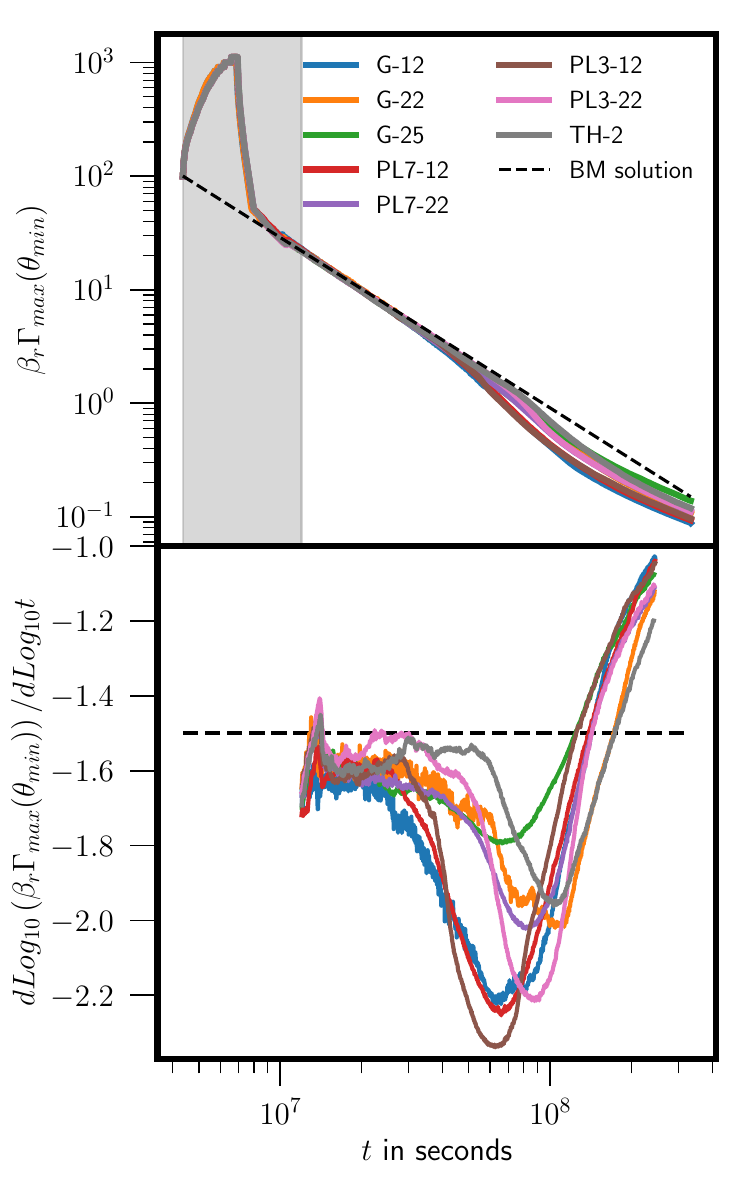}  
    \caption{Evolution of the maximum radial 4-velocity along the tip of the jet (i.e. the innermost simulation track) for different jet structures. Top panel shows four-velocity directly, bottom panel shows the slope of the four-velocity. The grey area marks an early transient feature prior to numerical convergence to the analytically expected evolution for the BM solution (both in slope and scale, and shown with a dashed line). The late time deviation of the different jets from the dashed profile is a consequence of further deceleration caused by jet spreading.}
    \label{fig:tip_Lorentz_factor}
\end{figure}

In Fig. \ref{fig:tip_Lorentz_factor}, we show the evolution of the radial four-velocity $\beta_{\rm tip} \gamma_{\rm tip}$ (in units of $c$) of the tip of the jet for different jet profiles. We also plot (with a black dashed line) the theoretically expected result from the BM solution given by,
\begin{equation}
\beta_{\textrm{tip}} \gamma_{\textrm{tip}} = 
\sqrt{\frac{17E_{\rm 0}}{16\pi\rho_{\rm ext}c^{5}}}t^{-\frac{3}{2}}.
\end{equation} 
We observe a transient numerical feature to appear at early times $t$, which subsequently fades out. Afterwards, the evolution can be seen to follow the theoretically expected behaviour. 
Such a transient feature arises as the mesh adapts to the initialized fluid variables, which contain sharp discontinuities and are subject to finite resolution. 
The convergence to the expected evolution once the feature fades indicates that the mesh and fluid have settled into a consistent coupled evolution.
At later times $t$, the evolution shows a prominent deviation from the usual power-law behaviour. This is a consequence of the sideways expansion of the jet. 

While the comparison to the BM solution serves as code verification, we can therefore already draw some first conclusions from Fig. \ref{fig:tip_Lorentz_factor} based on the post-break behaviour of the tip four-velocity. As expected, the smaller $\theta_{\rm c}$, the earlier the jet break occurs. Further, the transition is steeper for jets with sharper profiles (top-hat and steep power law) than for jets with smooth profiles (Gaussian, in particular if $\theta_{\rm w} > \theta_{\rm c}$). The impact of this smoothness on observations has been reported in the literature by e.g \cite{Lamb2021MNRAS} (an earlier discussion covering more theoretical work on the matter can be found in e.g. \citealt{Granot_2007}).

\begin{figure}
    \centering
    \includegraphics[width=\columnwidth]{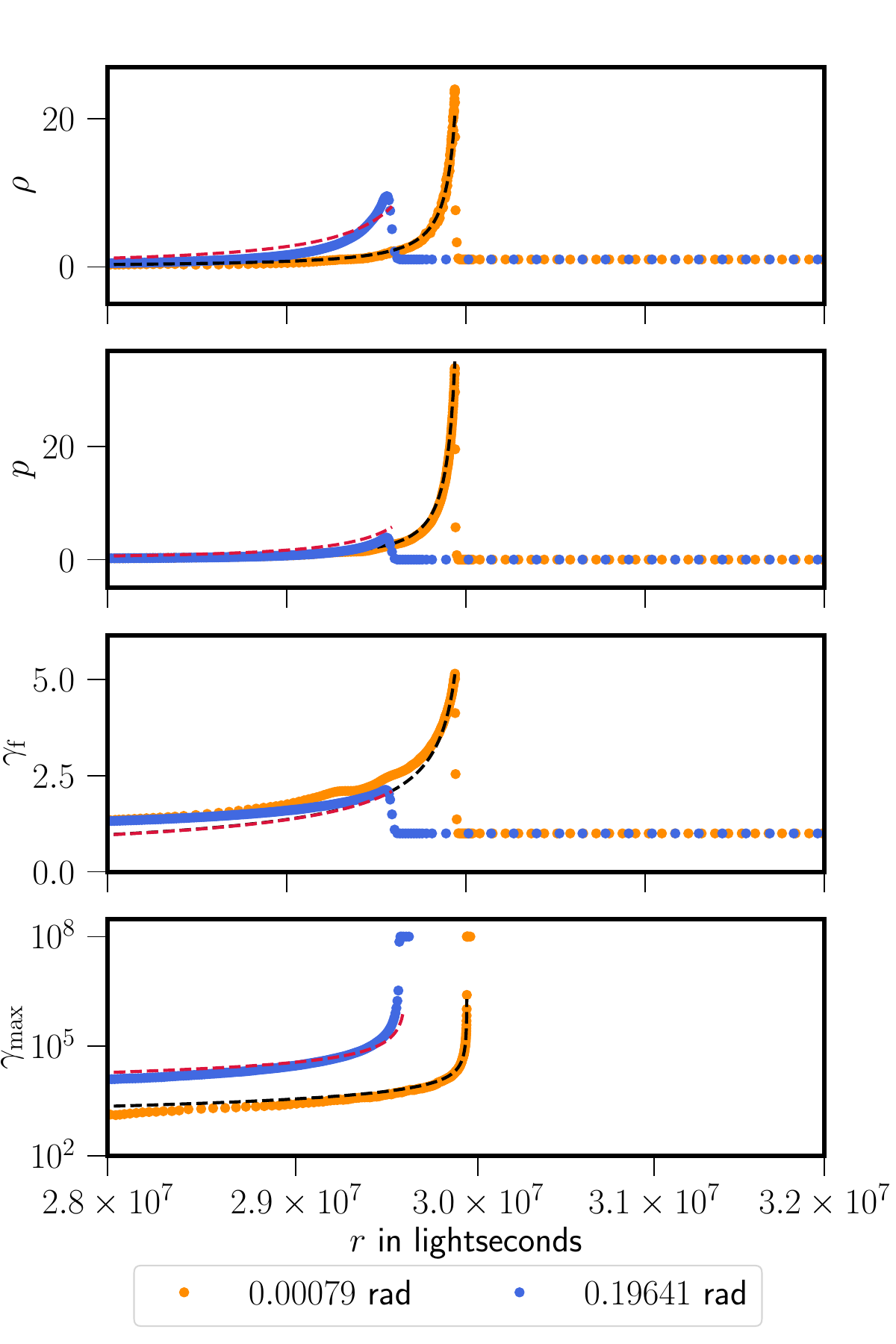}
    \caption{Radial profiles of (from top to bottom) density, pressure, fluid Lorentz factor, and $\gamma_{\rm max}$ of the non-thermal particle population experiencing synchrotron and adiabatic losses, in the shock downstream for Gaussian jet with $\theta_{\rm c}=0.1$ rad at $t_{\rm lab}=3\times 10^{7}$ s. Different coloured dots correspond to tracks with different angular coordinates; orange describes the innermost, and blue corresponds to the outermost track. Analytical behaviour, as described by \protect\cite{blandford_1976} and \protect\cite{GranotSari_2002}, for both angles is shown with coloured dashed curves; red corresponds to the innermost and black corresponds to the outermost track.}
    \label{fig:radial_gau}
\end{figure}

\begin{figure}
    \centering
    \includegraphics[width=\columnwidth]{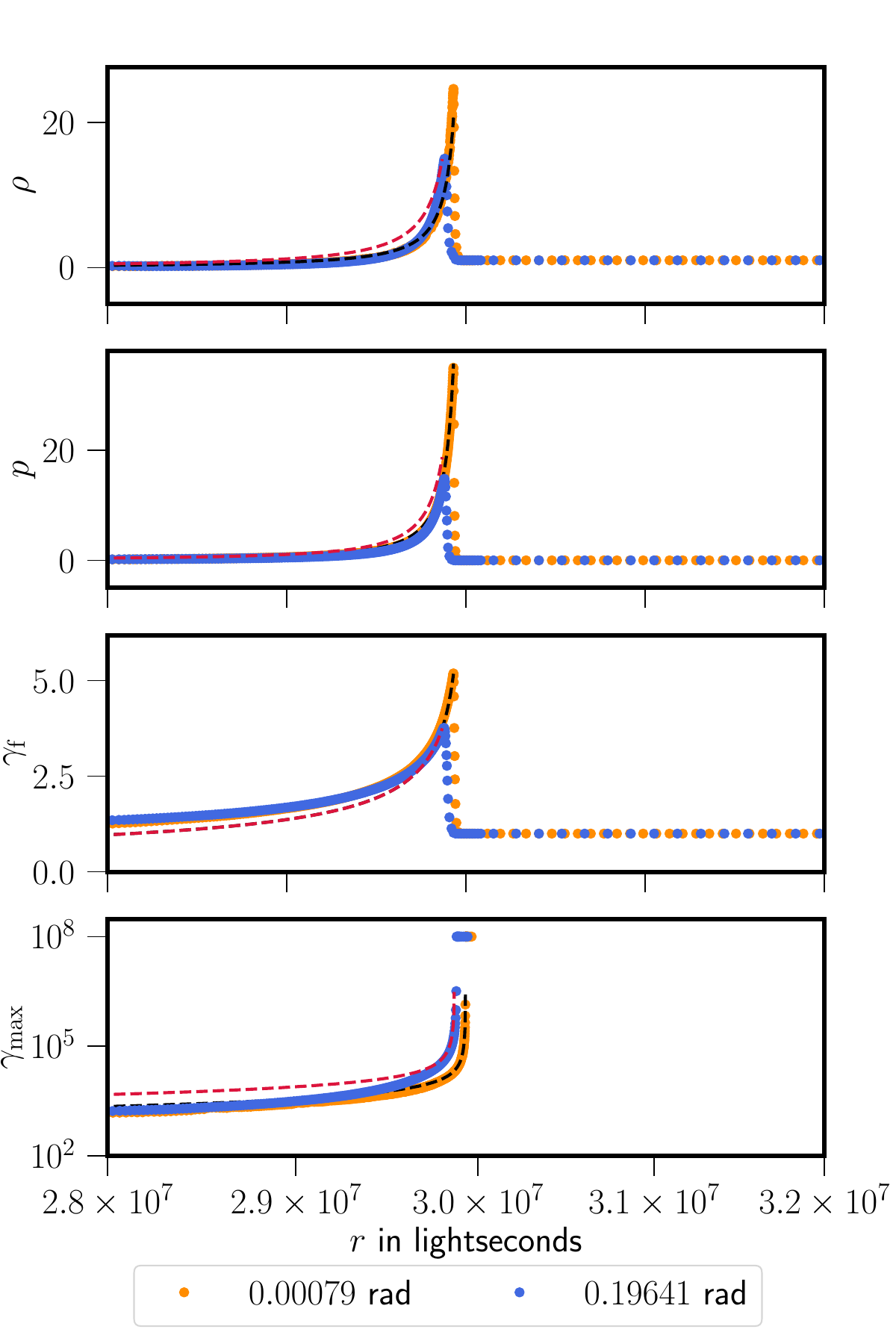}
    \caption{Same as Fig.~\ref{fig:radial_gau} but for a top hat jet.}
   \label{fig:radial_top}
\end{figure}

We further verify our simulations by directly testing the fluid profiles after the initial transient fades out against analytical solutions provided by \cite{blandford_1976} and \cite{GranotSari_2002}. As long as the outflow remains highly relativistic and as long as the impact of lateral motion remains minor, the simulated radial fluid profiles should match these self-similar solutions. In Figs.~\ref{fig:radial_gau} and \ref{fig:radial_top} we present the radial profiles of the fluid variables at $t=3\times 10^{7}$\,s, for the Gaussian and top hat jet structures respectively. In the top three panels we present the profiles of the fluid variables for two different angles. These correspond to the mid-points of the innermost (shown by orange dots) and outermost (shown by blue dots) radial tracks covered by the jet upon intialization. 
In the bottom panel of the figures, the radial profile of $\gamma_{\rm max}$ is shown. 
For both the angles the corresponding analytical profiles are shown with dashed lines.

The figures confirm that the density, pressure, fluid Lorentz factor and the non-thermal particles' $\gamma_{\rm max}$ follow the expected analytical profile for the innermost track, matching increasingly well with decreasing distance to the discontinuity at the shock front. The latter is expected for two reasons. On the one hand, the BM solution gets locally increasingly accurate with increasing fluid Lorentz factor. On the other hand, the further behind the shock front, the more likely the fluid profile is to be affected by any lingering impact from start-up transient features. We also observe, interestingly, that the outermost track also closely follows the analytical results, even though this is the track that is most exposed to the external medium. This can be understood from earlier results (e.g. \citealt{vanEerten_2012, vanEerten2013eConf}) indicating that full causal contact across the shock front is needed in order to drastically alter the fluid profile.

The asymptotic limit of the Sedov-Taylor regime of non-relativistic blast waves takes a long time to establish in practice (see e.g. \citealt{ZhangMacFadyen_2009} for a first numerically resolved multi-dimensional demonstration from simulation) and is therefore less relevant to a discussion of observable X-ray emission. In appendix \ref{appendix:wide_jets} we show an example of how our simulations eventually reach this limit dynamically.

\subsection{Jet spreading dynamics}
\label{section:jet_spreading}

\begin{figure*}
    \centering
    \includegraphics[width=\textwidth]{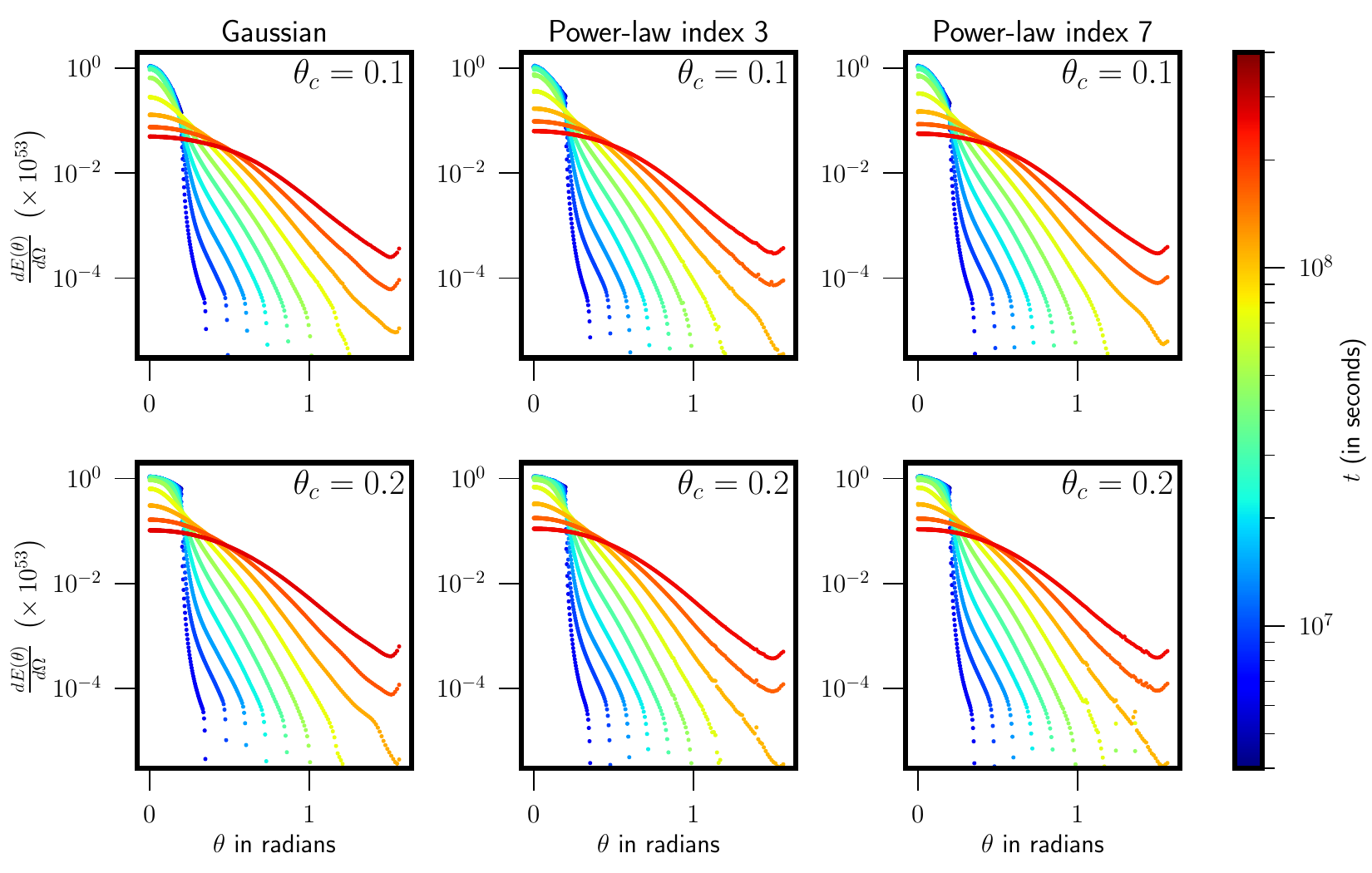}
    \caption{Temporal evolution of the lateral energy profile for different jet structures.
    \emph{Top panel} corresponds to jet structures having $\theta_{\rm c}=0.1$\,rad and \emph{bottom panel} shows the evolution for the structures with $\theta_{\rm c}=0.2$\,rad.
    Different colours of the curve correspond to different $t_{\rm lab}$ as shown in the colourbar.
    }
    \label{fig:gamma_evol}
\end{figure*}

\begin{figure*}
    \centering
    \includegraphics[width=\textwidth]{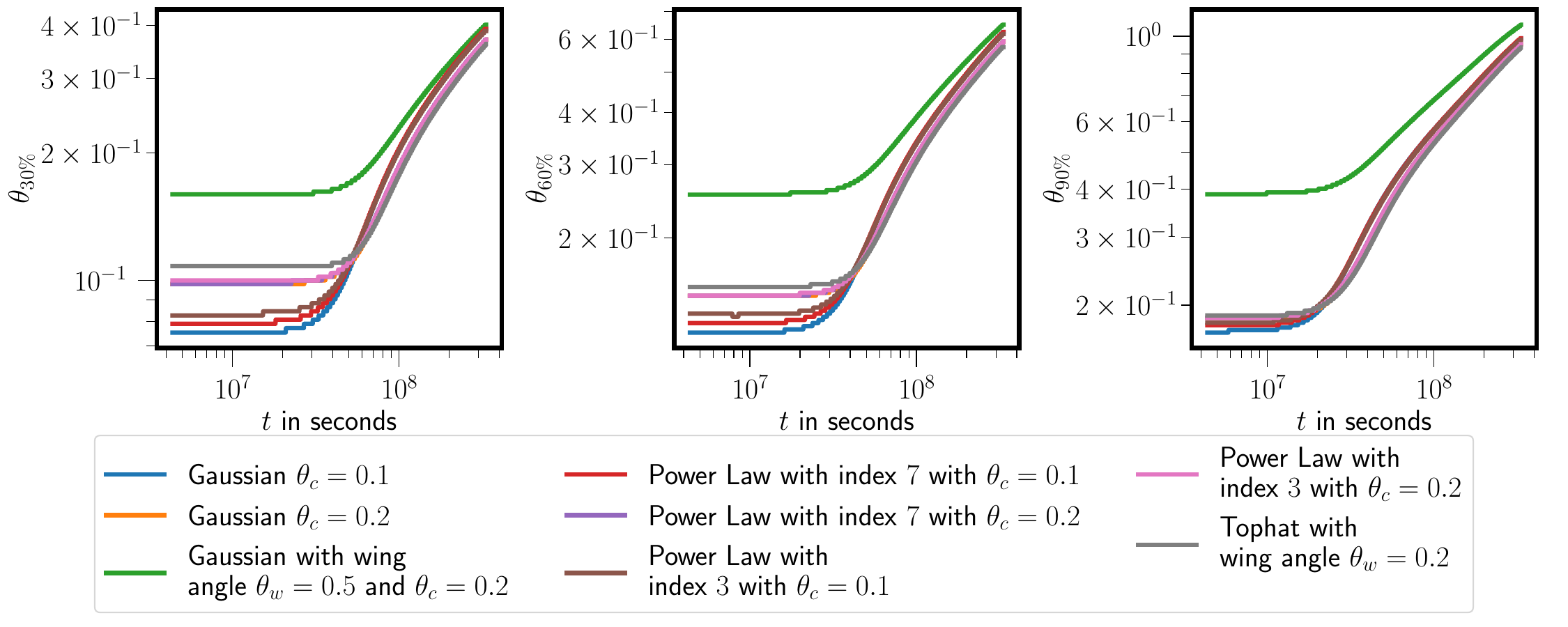}
    \caption{Evolution of $\theta$ containing $30\%$, $60\%$ and $90\%$ of the total jet energy for all the jet structures described in table~\ref{tab:1}.}
    \label{fig:theta_evol}
\end{figure*}

The simulations all show the expected lateral spreading behaviour after being initialized with only a radial velocity component. In Fig.~\ref{fig:gamma_evol} we show the evolution of the energy profiles of six simulations (Additional discussion of wide-jet case PL3-055 is referred to appendix \ref{appendix:wide_jets}). A sharp discontinuity can be seen to move sideways until it hits the equatorial plane at $\theta = \pi / 2$, at which point the jet encounters the counter-jet (i.e. a reflecting boundary condition) and energy piles up near the plane.

To have a more quantitative measure of the spreading behaviour in Fig~\ref{fig:theta_evol}, we show the evolution of $\theta$, containing $30\%$ (left), $60\%$ (middle), and $90\%$ (right) of the total jet energy (measured outward from the tip) to understand the spreading behaviour of different jet structures, extending to structured jets the top-hat jet analysis from e.g. \cite{vanEerten_2012, eerten_2012_boxfit, DeColle_2012, Granot_2012}. The smaller the cumulative energy cut-off, the longer the curves in the figure stay approximately flat. The key difference between top-hat jets and structured jets is that the latter start spreading sideways earlier on, as a consequence of being initialized with a lateral pressure gradient. The turn-over point to fast spreading is determined by the core angle $\theta_{\rm c}$. As in earlier work analyzing jets of similar width, the spreading does not fully enter into the exponential spreading regime expected for ultra-narrow jets \citep{vanEerten2013eConf, vanEerten_2018, Govreen-Segal_2024} that get to spend a sufficiently long time in a spreading regime (i.e. post onset of fast spreading) while still ultra-relativistic (i.e. before the tip Lorentz factor approaches trans-relativistic values). Once the jet has spread significantly beyond its initial opening angle, the memory of its structure upon initialization is effectively lost and all spreading profiles become similar in shape.

\begin{figure}
    \centering
    \includegraphics[width=\columnwidth]{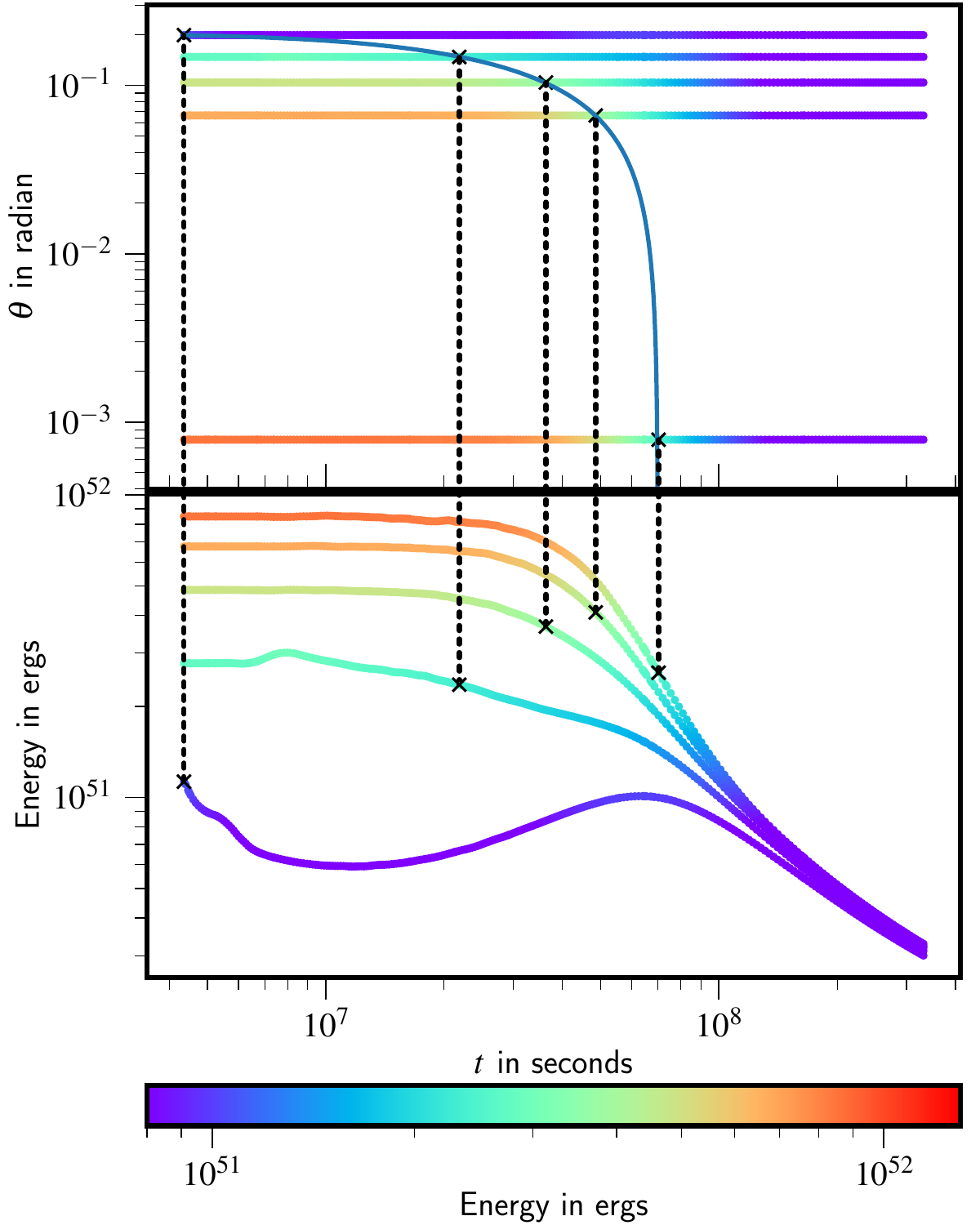}
    \caption{Comparison between the evolution of the lateral distribution of energy of a Gaussian structured jet (run G-12) and the inward motion of a rarefaction wave starting at the jet edge at the initial time of the simulation and moving inward with the speed of sound. The energy profile for five representative outflow angles is drawn from the simulation output, while the sound wave position along the jet front is calculated analytically. Crosses indicate the points where the sound wave crosses the plotted angle.}
    \label{fig:theta_track_Gaussian}
\end{figure}

\begin{figure}
    \centering
    \includegraphics[width=\columnwidth]{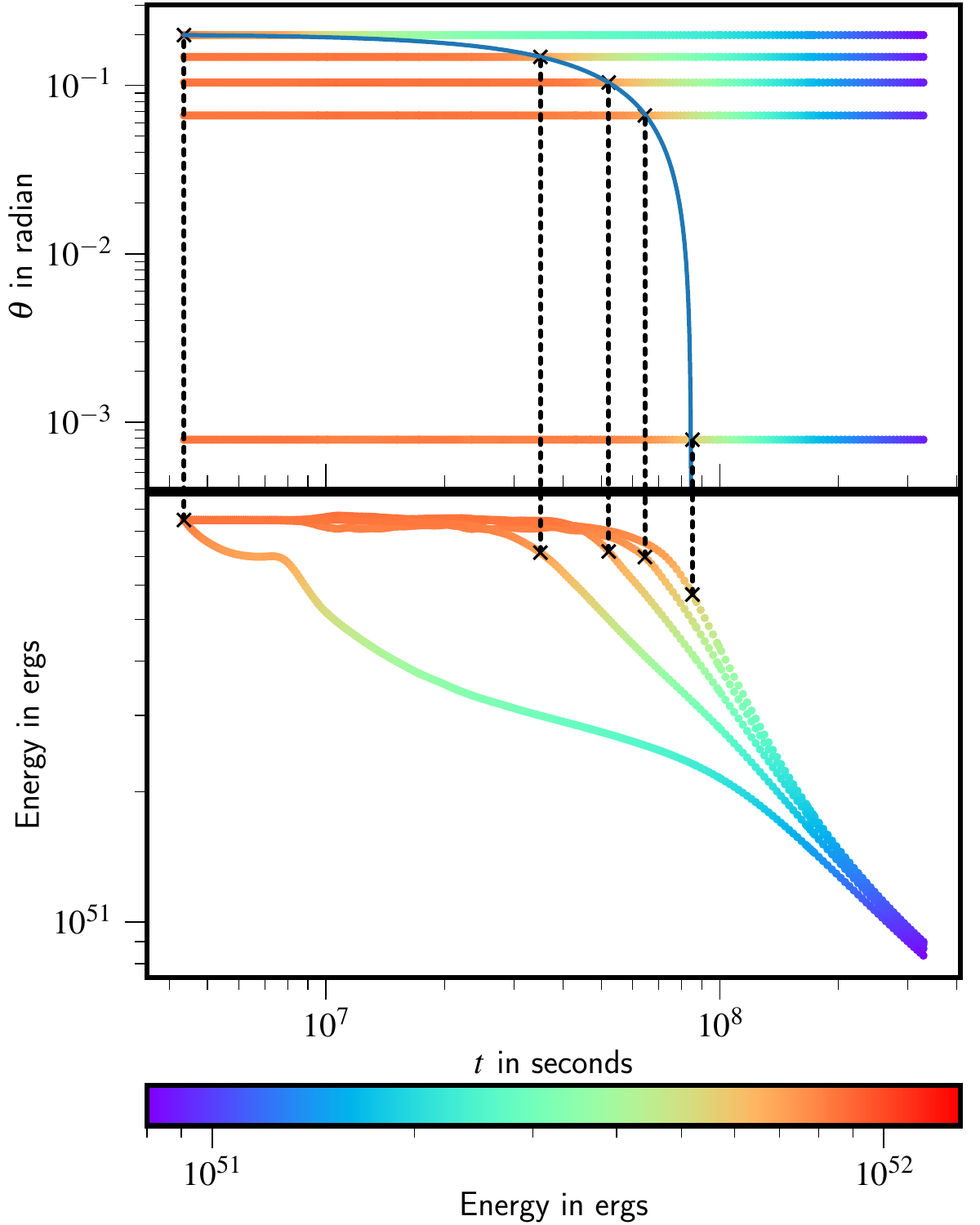}
    \caption{Same as Fig.~\ref{fig:theta_track_Gaussian} but for a top hat jet profile (TH-2).}
    \label{fig:theta_track_top}
\end{figure}

Jet spreading is known to be controlled by the establishment of causal contact between the outermost and the innermost parts of the jet \citep{vanEerten2013eConf}.
This causal contact is mediated by a rarefaction wave moving with the speed of sound (see e.g. \citealt{ZhangMacFadyen_2009, DeColle2012ApJdynamics} for numerical demonstrations of this feature). For an ultra-relativistic decelerating blast wave it can be shown that the angular position of the rarefaction wave obeys the following expression (\citealt{vanEerten2013eConf, vanEerten_2018}, see appendix \ref{appendix:sound_waves} for details):
\begin{equation}\label{eq:spread}
    \frac{d\theta}{dt} = -\frac{1}{2t\Gamma_{\rm S}},
\end{equation}
The negative sign on the right-hand side selects a sound wave solution moving inwards to lower angles. Prior to the passage of this sound wave, the flow at the shock front will remain radial (top-hat jets) or only spread in response to the initial lateral pressure gradient (structured jets).

In Figures \ref{fig:theta_track_Gaussian} and \ref{fig:theta_track_top} we show the impact of the lateral rarefaction wave for five different values of $\theta$ inside a Gaussian and a top-hat jet, including the outermost and the innermost tracks, depicted with horizontal lines in the top panel. 
The different colours of a single horizontal line depict the total integrated energy along a given angle as a function of time. 
The crossings of the solid blue curve across the horizontal lines mark the times when the perturbation hits a particular track (reaches a certain angular position). 
In the bottom panel, we show the evolution of the total energy of the tracks corresponding to the angles specified on the top panel.

We observe that for the top-hat jet there is very little energy change for angles not yet affected by the rarefaction wave along the shock front. However, because of the radial velocity profile, where at lower radii (and smaller outflow velocities) the rarefaction wave moving inwards is less suppressed by the Lorentz transformation to the lab frame, some minor motion is still seen. Further, both Figures show that the energy drops more steeply for angles further inwards once these are affected. This can be understood from the fact that lateral energy flux from closer to the tip has to pass through neighbouring angles first. It is only the tip that will never receive any energy from further inwards, and its energy drop after the onset of spreading is therefore the steepest.

The structured jet profile shows clear spreading already before the lateral rarefaction wave crosses a given angle. This is a direct consequence of the initial lateral pressure gradient.

\section{Results for radiation from structured Jets} 
\label{sec:rad}

The global and local cooling synchrotron prescriptions described in Section \ref{sec:emission} will lead to a characteristic SED of smoothly connected asymptotic power law regimes that has been well-established in GRB modelling and observations. 

Generally (assuming self-absorption to occur below any frequency of interest) in the slow cooling regime, the asymptotic limits of the SED can be written in terms of only $3$ parameters, $F_{\rm peak}$, $\nu_{\rm m}$ and $\nu_{\rm c}$ in the following manner (see e.g. \citealt{Sari_1998, Wijers_1999, granot_2002,eerten_2013_boosted}),
\begin{equation}\label{eq:global_flux}
    F_{\upnu} = \begin{cases} 
        F_{\textrm{peak}} \left( \frac{\nu}{\nu_{\rm m}} \right)^{1/3}, & \text{for } \nu < \nu_{\rm m} < \nu_{\rm c}, \\
        F_{\textrm{peak}} \left( \frac{\nu}{\nu_{\rm m}} \right)^{\frac{1-p}{2}}, & \text{for } \nu_{\rm m} < \nu < \nu_{\rm c}, \\
        F_{\text{peak}} \left( \frac{\nu_{\rm c}}{\nu_{\rm m}} \right)^{\frac{1-p}{2}} \left( \frac{\nu}{\nu_{\rm c}} \right)^{-\frac{p}{2}}, & \text{for } \nu_{\rm m} < \nu_{\rm c} < \nu.
    \end{cases}
\end{equation}

In the fast cooling regime, we have instead
\begin{equation}\label{eq:global_flux_fast}
    F_{\upnu} = \begin{cases} 
        F_{\text{peak}} \left( \frac{\nu}{\nu_{\rm c}} \right)^{1/3}, & \text{for } \nu < \nu_{\rm c} < \nu_{\rm m}, \\
        F_{\text{peak}} \left( \frac{\nu}{\nu_{\rm c}} \right)^{-\frac{1}{2}}, & \text{for } \nu_{\rm c} < \nu < \nu_{\rm m}, \\
        F_{\text{peak}} \left( \frac{\nu_{\rm m}}{\nu_{\rm c}} \right)^{-\frac{1}{2}} \left( \frac{\nu}{\nu_{\rm m}} \right)^{-\frac{p}{2}}, & \text{for } \nu_{\rm c} < \nu_{\rm m} < \nu.
    \end{cases}
\end{equation}

We now proceed to investigate how the evolution of these three parameters is affected by the different jet structures and approaches to electron cooling.

\subsection{Shape and evolution of spectral features}
\label{sec:spectral_features}

\begin{figure*}
    \centering
    \includegraphics[width=\textwidth]{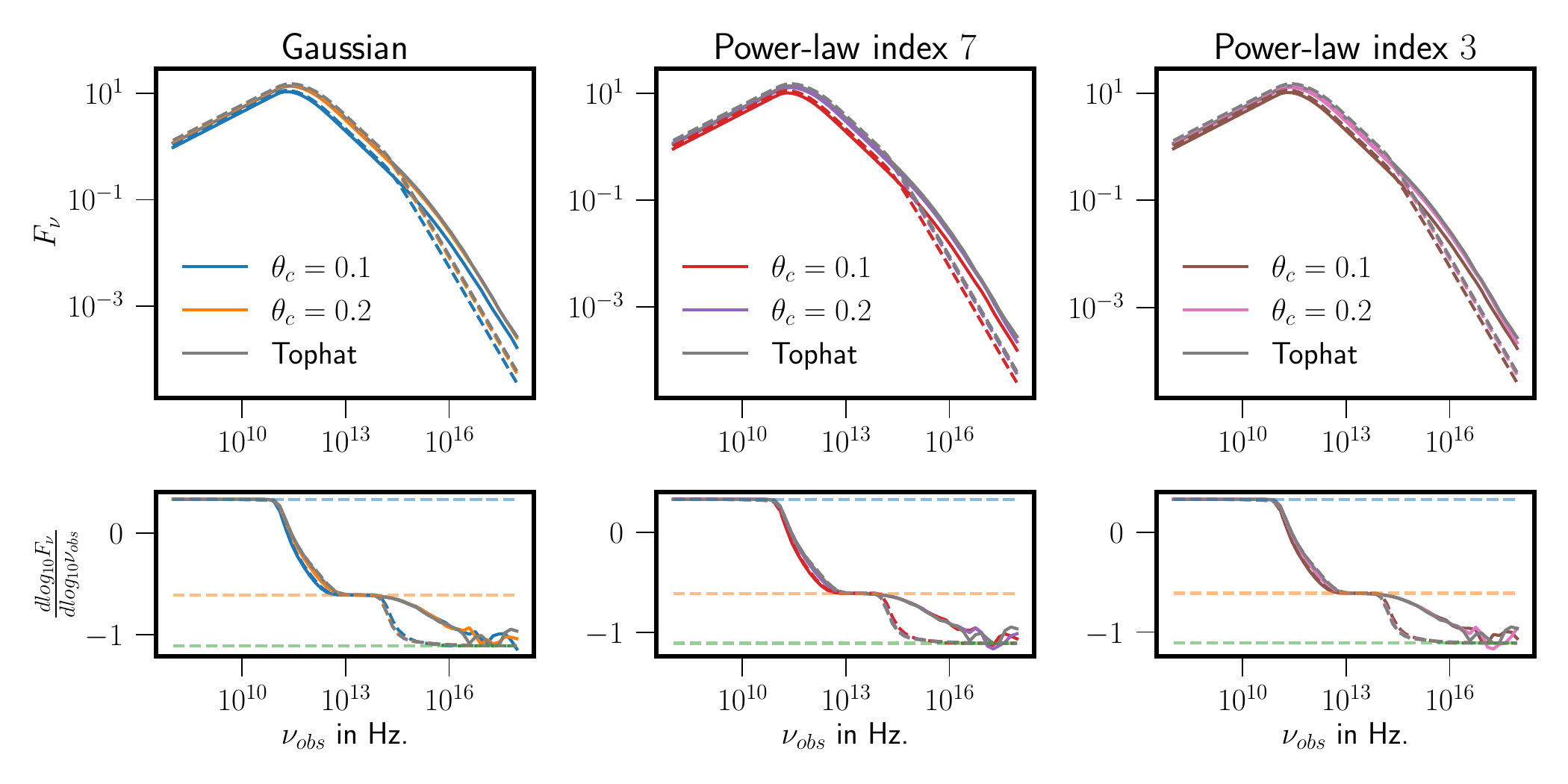}
    \caption{\emph{Top:} On-axis SED at $t_{\rm obs}=10^{5}$\,s for different jet structures (with $\theta_{\rm w}=0.2$). 
    Solid lines correspond to light curves computed from the local cooling approach, and dashed lines correspond to the global cooling approach.
    The SED of the Tophat jet computed from both the cooling approaches is presented with the black coloured lines for comparison. 
    \emph{Bottom:} The slope of the SED is shown for different jet structures. 
    Dashed curves present the global cooling SEDs, while the dotted curve corresponds to local cooling SEDs.
    Further, the slopes for $3$ different spectral regimes are shown by dashed horizontal lines; blue corresponds to $1/3$, orange corresponds to $(1-p)/2 = -0.61$, and green to $-p/2 = -1.11$. The SEDs are all in the slow-cooling regime.
    }
    \label{fig:SED_on_axis}
\end{figure*}

Fig.~\ref{fig:SED_on_axis} presents the on-axis SEDs for different jet structures under both cooling prescriptions, with the Tophat jet shown as a black line for comparison.
The solid lines correspond to the SED based on local cooling, whereas the dashed line represents the SED computed using the global cooling prescription.
The bottom panel displays the SED slopes, revealing three distinct regimes. 
The slope transitions from $1/3$ at low frequencies to $(1-p)/2$ at $\nu_{\rm m}$, then to $-p/2$ at $\nu_{\rm c}$, where in this work we assume $p=2.22$ due to relativistic shock acceleration. The match to intermediate slope $(1-p)/2 = -0.61$ (rather than $-1/2$) implies that we are in the slow cooling regime for all the jet structures shown in the Figure.

In Fig. \ref{fig:SED_on_axis} the differences between local and global cooling stand out clearly, regardless of jet structure. In a local cooling approach, the cooling break is shifted to a substantially higher frequency (consistent with the top-hat results reported in \citealt{Ayache_2022}). Furthermore, the bottom panels showing the slopes of the SEDs confirm that the cooling break transition is far more smooth in a local cooling approximation.

To examine the shape and evolution of the SEDs in more detail, we use a combined smooth power-law fit function similar to those used by \cite{GranotSari_2002} to extract their characteristic features:
\begin{align}
    F_{\upnu}(\nu)
 = & \; F_{\rm peak}\,
\left[\,\left(\tfrac{\nu}{\nu_{\rm m}}\right)^{-s\beta_{\rm 1}}
      + \left(\tfrac{\nu}{\nu_{\rm m}}\right)^{-s\beta_{\rm 2}}\,\right]^{-\frac{1}{s}} \times \nonumber \\
&  \left[\,1 + \left(\tfrac{\nu}{\nu_{\rm c}}\right)^{\,s_{\rm c}(\beta_{\rm 2}-\beta_{\rm 3})}\right]^{-\frac{1}{s_{\rm c}}},
\end{align}
where $F_{\rm peak}$ a scale factor, $\beta_{\rm 1}$, $\beta_{\rm 2}$ and $\beta_{\rm 3}$ asymptotic power-law slopes and $s$ and $s_{\rm c}$ the sharpness parameters for the power-law transitions across $\nu_{\rm m}$ and $\nu_{\rm c}$ respectively. The slopes are kept fixed at their theoretically expected values, i.e. $\beta_{\rm 1} = 1/3$, $\beta_{\rm 2} = (1-p)/2$, $\beta_{\rm 3}= - p/2$. For completeness we note that we fit this function to numerically generated SED curves with in practice around 60 logarithmically spaced frequencies from $10^{8}$ Hz to $5\times 10^{17}$ Hz, for 27 observer times starting from $t_{\rm obs}=2\times10^{4}$ s to $t_{\rm obs}=2\times10^{7}$ s.

\begin{figure*}
    \centering
    \includegraphics[width=\textwidth]{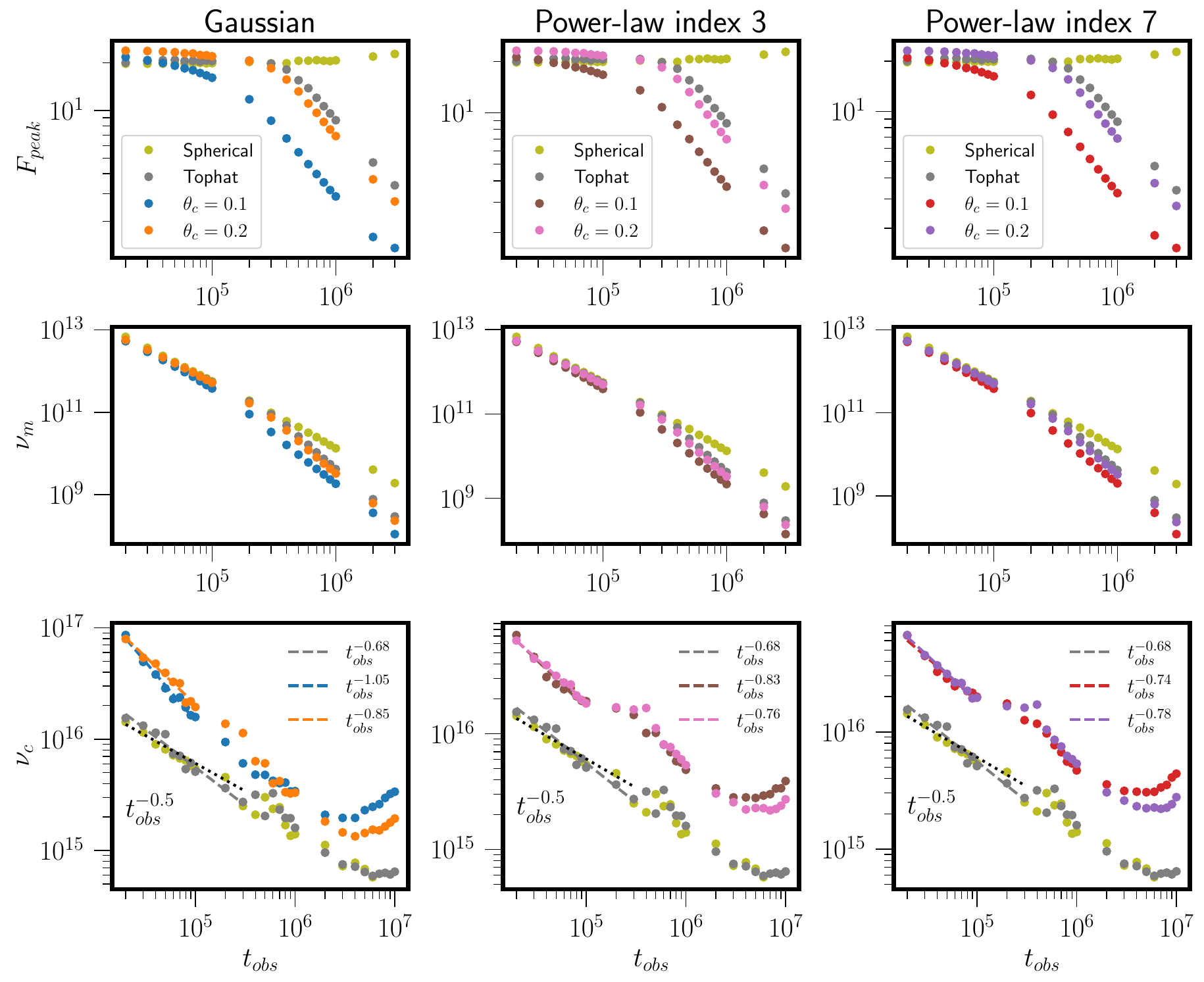}
    \caption{Evolution peak flux $F_{\rm peak}$, injection break $\nu_{\rm m}$ and cooling break $\nu_{\rm c}$, shown in the top, middle, and bottom panels, respectively, for different jet structures in the local cooling scenario. The expected evolution $\nu_{\rm c} \propto t_{\rm obs}^{-1/2}$ in the top-hat case is shown with black dotted curves for comparison.}
    \label{fig:cool_evol_loc}
\end{figure*}

In Fig.~\ref{fig:cool_evol_loc}, we present the evolution with time of the characteristic features for various structures in the local cooling scenario, alongside output for a perfectly spherical blast wave for comparison.
The top panel of the figure shows the evolution of $F_{\rm peak}$ for different jet structures, the middle panel shows the evolution of $\nu_{\rm m}$, and the bottom panel demonstrate the evolution of $\nu_{\rm c}$. 
In the bottom panel, we also show power-law fits to the data points up to $t_{\rm obs}=10^{5}$ s for structured jets up to $4 \times 10^5$ s for the top-hat jet (i.e. the time interval before the occurrence of jet-break) to quantify the evolution of $\nu_{\rm c}$.

The evolution of $F_{\rm peak}$ can be seen to change more smoothly around the jet break (occurring roughly within the range $1 \times 10^5 - 4 \times 10^5$ s) for structured jets than for a top-hat jet, as expected.
The evolution of $\nu_{\rm m}$ is largely similar between the top-hat jet and structured jets with core width comparable to the total width of the top-hat comparison jet. Narrower structured jet profiles show an earlier onset of the drop in $\nu_{\rm m}$ than the top-hat jet shows.
As expected, in the absence of a jet break the spherical blast wave shows neither change in behaviour over time for $F_{\rm peak}$ nor for the critical frequencies.

Most significant is the difference in cooling break $\nu_{\rm c}$ between the top-hat jet and the structured jets, with the latter showing both a large upward shift and a difference in temporal slope. With regards to temporal slope, it is the Gaussian jet that stands out most. This steep slope is surprisingly close to e.g. the observationally reported result by \cite{filgas_2011}, although we caution against directly comparing two numerical values for the temporal slope obtained in different contexts and through non-identical means. Furthermore, the top-hat fit result is steeper than the expected value ($\nu_{\rm c} \propto t_{\rm obs}^{-1/2}$, shown with a black dotted line; a closer numerical match was previously reported by \citealt{Ayache_2022}), so while the difference in estimated temporal slope between top-hat and structured jets is noticeable, we caution that this signifies  both significant numerical noise and a limited range for the pre-jet asymptotic limit to reveal itself.
Both of these imply that the best-fit slope values shown should be assumed to come with substantial error bars even while they show a persistent trend of increasing temporal slope from top-hat to structured profiles. For the top-hat result the numerical error would be about 0.1, based on the scatter in the $\nu_{\rm c}$ values around the best fit, and about 0.18 based on the difference between the measured slope and the analytically expected value (to the extent that this difference is not a consequence of the lack of time spent in the analytically expected regime).

\begin{figure*}
    \centering
    \includegraphics[width=\textwidth]{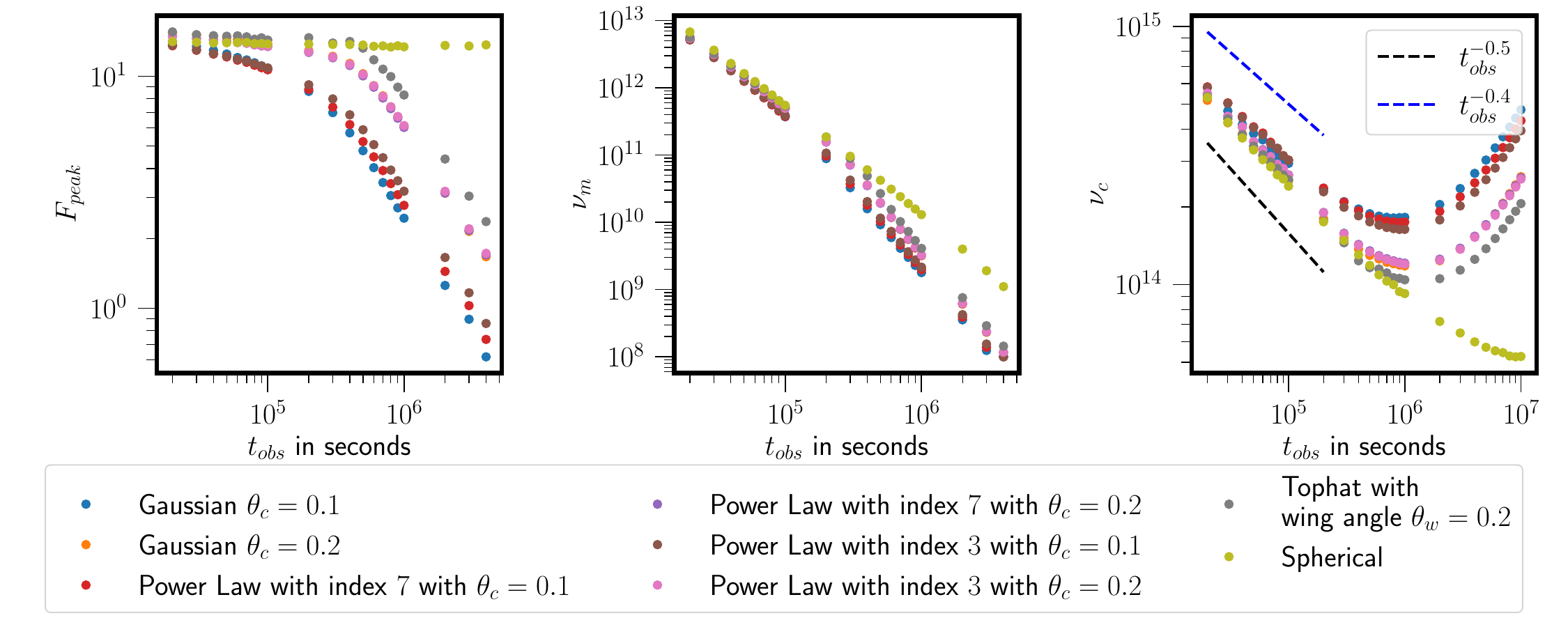}
    \caption{Evolution $F_{\rm peak}$, $\nu_{\rm m}$ and $\nu_{\rm c}$, now assuming global cooling.}
    \label{fig:cool_evol_glob}
\end{figure*}

In Fig.~\ref{fig:cool_evol_glob} we show the evolution of the characteristics of the SED assuming global cooling. The top-hat case can be seen to be consistent with \cite{eerten_2013_boosted}. We also see a similar evolutionary trend to local cooling for $F_{\rm peak}$ and $\nu_{\rm m}$. 
On the other hand, for $\nu_{\rm c}$, we see a general trend markedly different from local cooling for different structures, where an initial power-law decrease is followed by a turnover and rise, consistent with global cooling results from both \cite{eerten_2013_boosted} and \cite{Ayache_2022}.
We further note that $\nu_{\rm c}$ for global cooling at all times and for all structures stays below $\nu_{\rm c}$ observed for local cooling. 

\begin{figure*}
    \centering
    \includegraphics[width=\textwidth]{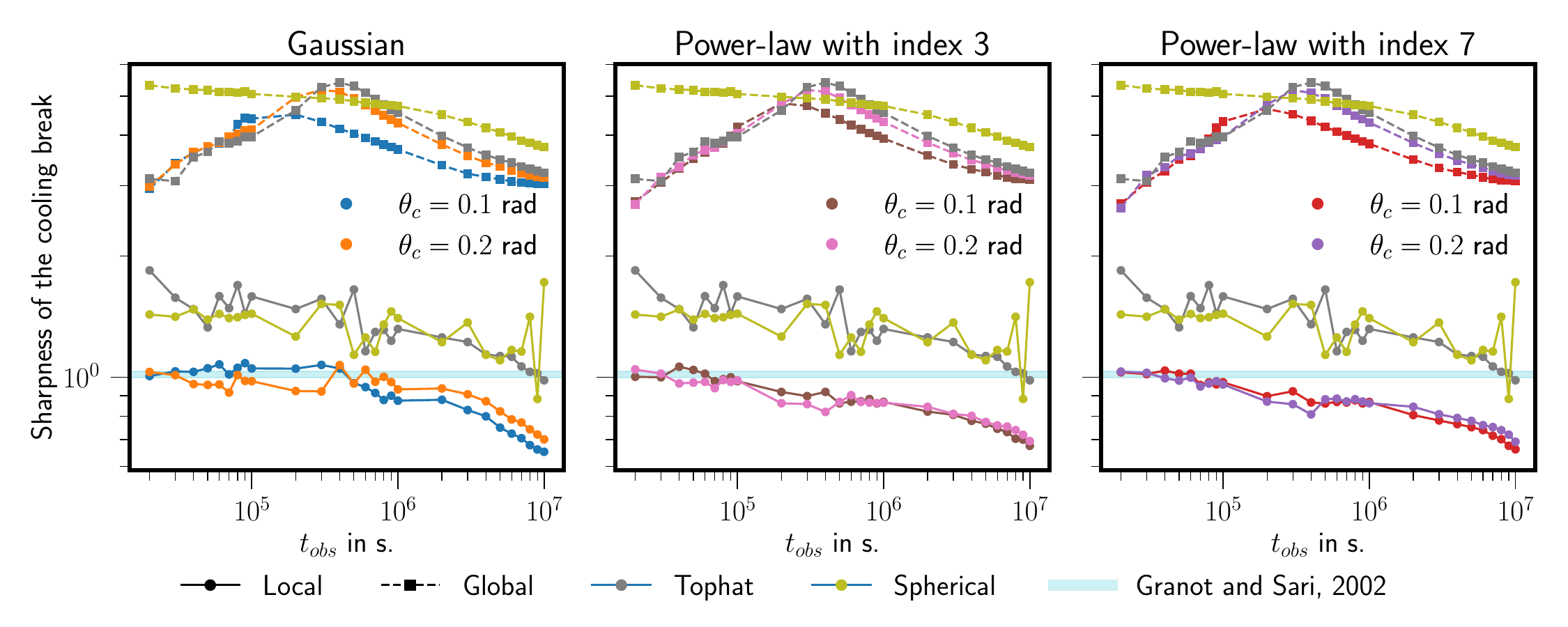}
    \caption{Evolution of the sharpness of the cooling break for different structures in both the cooling scenarios. 
    The cooling breaks obtained from the global cooling approximation are shown with squares, while the cooling breaks using local cooling are shown with dots. 
    The corresponding evolution for the top-hat jet is shown in black for comparison.
    }
    \label{fig:cool_evol_sharp}
\end{figure*}

Another interesting feature for follow-up from Fig.~\ref{fig:SED_on_axis} is the sharpness of the transition of the spectral slope from $(1-p)/2$ to $-p/2$ for the local cooling scenario shown in the bottom panel.
To study this sharpness quantitatively, we plot the sharpness parameter $s_{\rm c}$ for different structures %\textcolor{blue}{alongwith the spherical blast wave and for} 
and different cooling prescriptions at various $t_{\rm obs}$. Again, results for a spherical blast wave are included for comparison. We show the plot in Fig. ~\ref{fig:cool_evol_sharp}.

The sharpness parameter can be seen to be lower for local cooling than for global cooling for all jet structures, implying a smoother spectral transition around $\nu_{\rm c}$ for local cooling SEDs, in line with our earlier results. 
For the top-hat jet, the local cooling sharpness parameter has a value between $1$ and $2$, consistent with \cite{GranotSari_2002} shown in light blue in the Figure. The remaining discrepancy between our results and theirs is likely due to us applying a broken power-law spectrum for the synchrotron emission. By contrast, the calculation in \cite{GranotSari_2002} employed a full analytical synchrotron spectrum, which naturally produces a smoother transition than our adaptation of broken power-law profiles.

For global cooling SEDs, the sharpness of the cooling break can be seen to peak around the time of the jet break, going from increasing sharp spectral breaks back to smoother breaks. An interesting artifact from the global cooling approach is the shearing layer along the edge of the jet already impacting the spectral sharpness of the cooling break prior to the nominal jet break in the light curve. This is because the slower-moving material of this layer will produce emission that is less beamed than faster-moving material within the jet. The impact can clearly be seen in the early-time difference between sphere and jet models in the Figure. We discuss this further in Appendix \ref{appendix:shearing_layer}.

\begin{figure*}
    \centering
    \includegraphics[width=\textwidth]{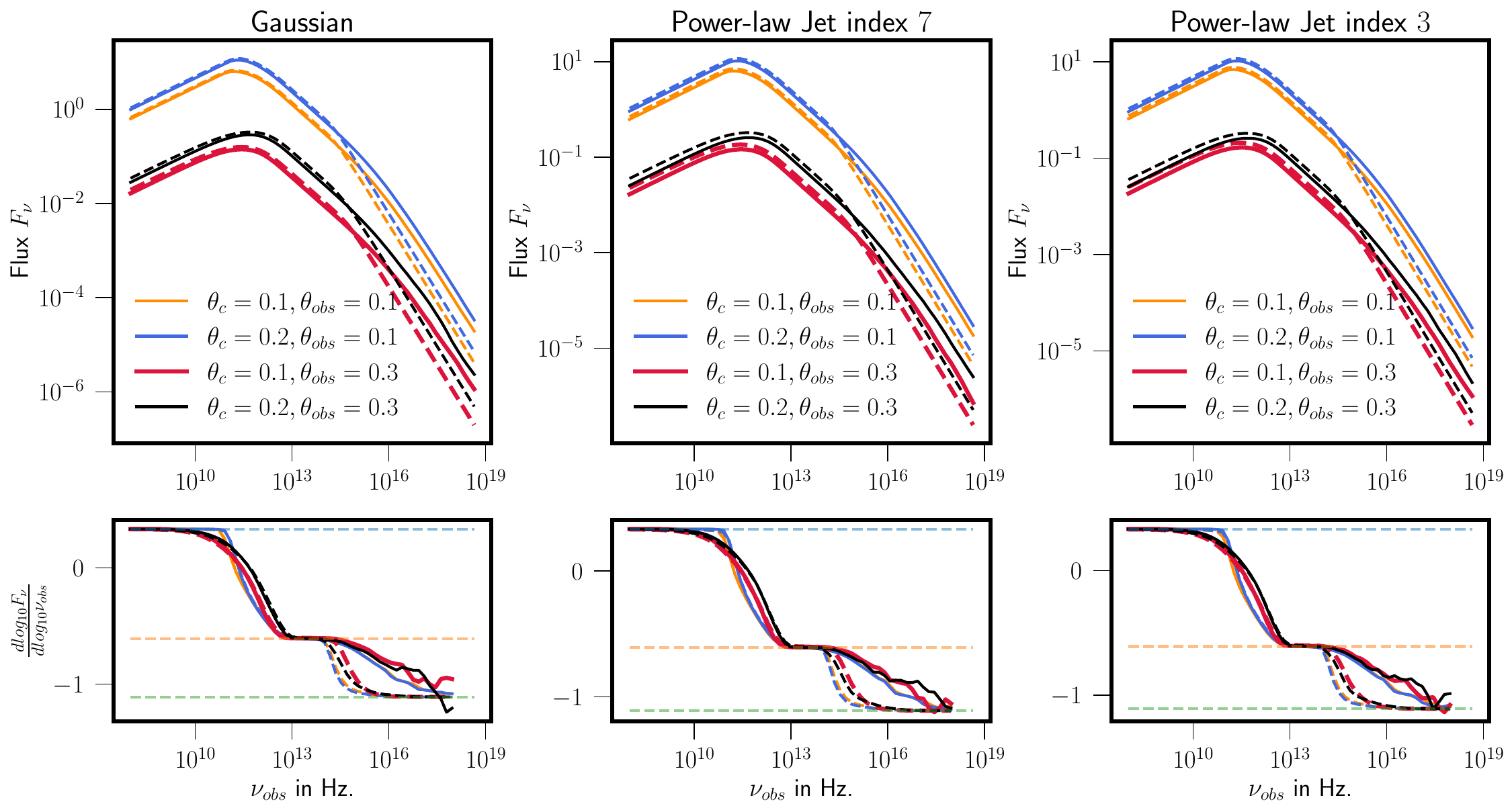}
    \caption{\emph{Top panel:} Off-axis SED at $t_{\textrm{obs}}=10^{5}$\,s for different jet structures using both the cooling prescription for two different observing angles, $0.1$ and $0.3$\, rad.
    The SED from the local cooling approach is shown with the solid lines, while the SEDs from the global cooling approach are shown with dashed lines.
    \emph{Bottom Panel:} Shows the slopes of the SEDs shown in the top panel.  
    The slopes for $3$ different spectral regimes are shown by dashed horizontal lines; blue corresponds to $1/3$, orange corresponds to $(1-p)/2 = -0.61$, and the green line shows the slope of $-p/2 = -1.11$.}
    \label{fig:off_sed}
\end{figure*}

Finally, in Fig.\ref{fig:off_sed} we present the off-axis SED for different jet structures considering both cooling scenarios. 
The figure shows the SED for two different observer angles $0.1$ and $0.3$\,rad.
The flux for observing angle $0.3$\,rad shows a significant decrease as compared to the other observing angle. 
The bottom panel illustrates the spectral slopes. 
Here, too, we see the difference between local and global cooling, as in Fig. \ref{fig:SED_on_axis}. 

\subsection{The emission region}

\begin{figure}
    \centering
    \includegraphics[width=\columnwidth]{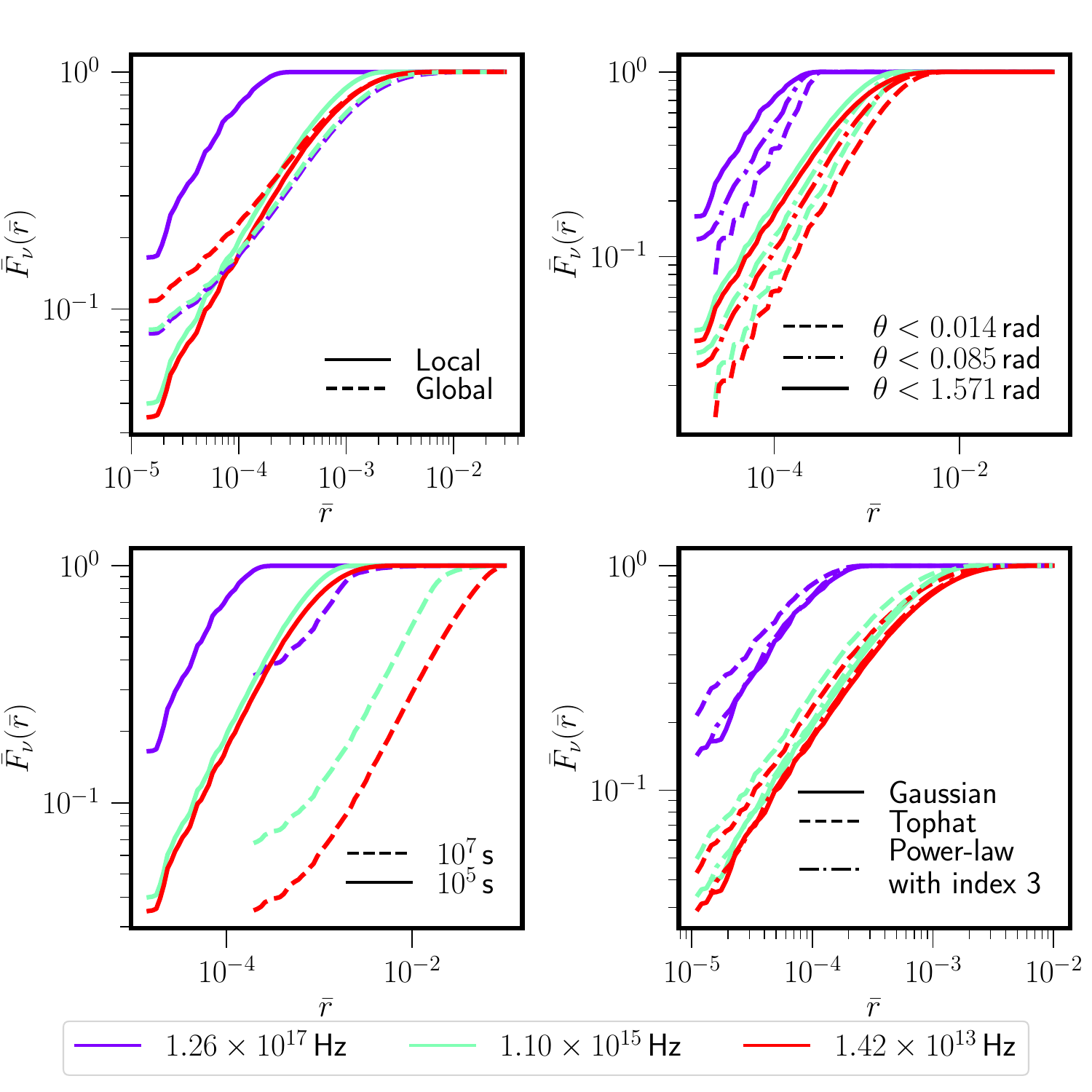}
    \caption{Normalized cumulative flux as function of scaled distance to the shock front, for three representative frequencies, mapping the extent of the region behind the shock front that contributes to the observed flux (with the cumulative flux for higher frequency emission reaching its peak closer to the shock front). Unless stated otherwise in the legend or below, all curves refer to measurements taken at $t_{\rm obs} = 10^5$\,s and to Gaussian jet simulation G-12. \emph{Top left} shows the difference between a local and global cooling approach. \emph{Top right} shows how cumulative flux changes when only outflow within a cut-off angle $\theta$ is included. \emph{Bottom left} shows a comparison between cumulative flux regions across two observer times. \emph{Bottom right} shows a comparison between jet structures (Gaussian G-12, Top hat TH-2, Power Law PL3-12).}
    \label{fig:rbar_loc}
\end{figure}

Local cooling provides insight into the particle population in the downstream region of the forward shock, as it accounts for the collective contributions of all particle distribution functions behind the shock when computing the radiation.
Consequently, it helps to quantitatively estimate the width of the region behind the shock responsible for emission at a certain frequency. We define a normalized distance $\bar{r} (\theta, t, r)$ from the shock front according to
\begin{equation}
\bar{r} \left( \theta, t, r \right) = \frac{R_{\rm s}\left( \theta, t \right) - r}{R_{\rm s}\left( \theta, t \right) - r_{\rm min}}
\label{eq:r_bar}
\end{equation}
Here $r_{\rm min}$ is the minimum radius of the computational grid, kept fixed for all simulations. The shock front radius $R_{\rm s}$ is (numerically) taken to be the radius having the maximum 4-velocity for a given track. This definition identifies $\bar{r}=0$ with the shock front and $\bar{r}=1$ with the minimum boundary of the computational grid.

For a fixed value of $t_{\rm obs}$, we can use Eq. \ref{eq:EDS} to eliminate the dependence on $t$ in Eq. \ref{eq:r_bar}, i.e. by using
\begin{equation}
t\left( r, \theta, \phi \right) = t_{\rm obs} + \frac{r \mu \left( \theta, \phi \right)}{c}.
\end{equation}
We can now define a cumulative flux $\bar{F}_{\upnu}$ according to
\begin{equation}
\bar{F}_{\upnu} \left( \bar{r} \right) \equiv \frac{ \int_{\theta=0}^\pi \int_{\phi=0}^{2\pi} \int_{\bar{r}'=0}^{\bar{r}} \frac{d F_{\upnu}}{d \bar{r}' d \theta d \phi} d \bar{r}' d \theta d \phi}{\int_{\theta=0}^\pi \int_{\phi=0}^{2\pi} \int_{\bar{r}'=0}^{1} \frac{d F_{\upnu}}{d \bar{r}' d \theta d \phi} d \bar{r}' d \theta d \phi}.
\label{eq:cum_flux}
\end{equation}
Note that the primes applied to $\bar{r}$ in the integrals of Eq. \ref{eq:cum_flux} are merely to indicate where it is used an an integration variable, and do not imply anything about reference frames (i.e. all radii are in the lab frame).

In Fig.~\ref{fig:rbar_loc} we show how $\bar{F}_{\upnu}(\bar{r})$ depends on $\bar{r}$ for different frequencies.
By construction, all curves in the figure increase monotonically with $\bar{r}$ and saturate at $1$.
The value of $\bar{r}$ at which the curve saturates gives a scale length behind the shock up to which the particles responsible for emission at that frequency reside. The expectation is that (certainly in a local cooling approach) $\bar{F}_{\upnu}$ will saturate at smaller $\bar{r}$ for higher frequencies, since the higher the frequency of the emission, the hotter the electron population needs to be to produce this emission.

In the top-left panel we show how $\bar{F}_{\upnu}$ differs between frequencies and approaches to cooling. The global cooling curves all cluster together regardless of frequency, even if some difference is seen between $1.26 \times 10^{17}$ Hz (above the cooling break, see Fig. \ref{fig:SED_on_axis}) and the others, while the curve for $1.10 \times 10^{15}$ Hz lies (very) slightly above $1.42 \times 10^{13}$ Hz. For local cooling, instead, the cooling break has a strong impact, confirming that the size of the region producing the emission is an order of magnitude smaller above the cooling break.

In the top-right panel we show the impact on $\bar{F}_{\upnu}$ when emission within a truncation angle $\theta$ is considered. Solid curves correspond to the full grid, dash-dotted curves to emission from $\theta < 0.085$ rad (inner 50 simulation tracks) and dashed curves to $\theta < 0.014$ rad (inner 10 simulation tracks). There are two competing effects at play that impact the extent of the emission region as a function of angle. On the one hand, for a given observer time, emission from larger angles will originate earlier than emission from closer to the tip of the jet. This implies that a wider emission region is expected when a smaller truncation angle $\theta$ is included, since early emission times correspond to a thinner blast wave. This effect will be present even for top-hat jets. On the other hand, jet structure will alleviate this difference, depending on how steeply the energy drops off with angle, because the lower the energy of the jet, the wider the (post-deceleration stage) blast wave will be compared to a more energetic region. Note that for observer frequencies above $\nu_{\rm m}$ we expect the image of the afterglow on the sky to be limb-brightened \citep{Granot_1999, vanEerten_2010}, which implies that the emission region width at the jet tip will be less relevant than that at larger values of $\theta$ for the purpose of assessing where the emission seen at a given observer time is produced.

In the bottom-left panel we compare between the observer time $10^5$ s we have used so far and a later observer time of $10^7$ s. The same pattern persists, where higher frequencies probe a more narrow region behind the shock than lower frequencies. However, at later times, the blast wave will become altogether wider and an overall shift to a larger saturation value of $\bar{r}$ is expected. Furthermore, for simulation $G-12$ shown here, the difference between $1.10 \times 10^{15}$ Hz and $1.42 \times 10^{13}$ Hz stands out more clearly at $10^7$ s. This can be understood from the bottom left panel of Fig. \ref{fig:cool_evol_loc}, where the cooling break $\nu_{\rm c}$ can be seen to be significantly closer to $1.1 \times 10^{15}$ s. Because the turnover in the spectrum around the cooling break is smooth (as demonstrated in the bottom panels of Fig. \ref{fig:SED_on_axis}), this will directly affect the width of the emission region.

In the bottom-right panel, we show how $\bar{F}_{\upnu}$ varies between jet structures. It can be seen that the increase in $\bar{F}_{\upnu}(\bar{r})$ with $\bar{r}$ is steeper for power-law and Gaussian jets than for for the Tophat jet. This is consistent with our earlier statement about the jet observations showing limb-brightening images if they are spatially resolved. The curves for the power-law profile lying in between the top-hat and Gaussian jet cases is consistent with the power-law expression for the energy profile being an intermediate case between both extremes.

In summary, the conclusions drawn from Fig. \ref{fig:rbar_loc} demonstrate two things: one, the global cooling approach lacks the ability to capture the expected spatial distribution of different particle populations behind the shock; two, the angular distribution of the widths of the emission regions will differ between jet structures.

\subsection{(X-ray) Light curves}
\label{sec:light_curves}

\begin{figure*}
    \centering
    \includegraphics[width=\textwidth]{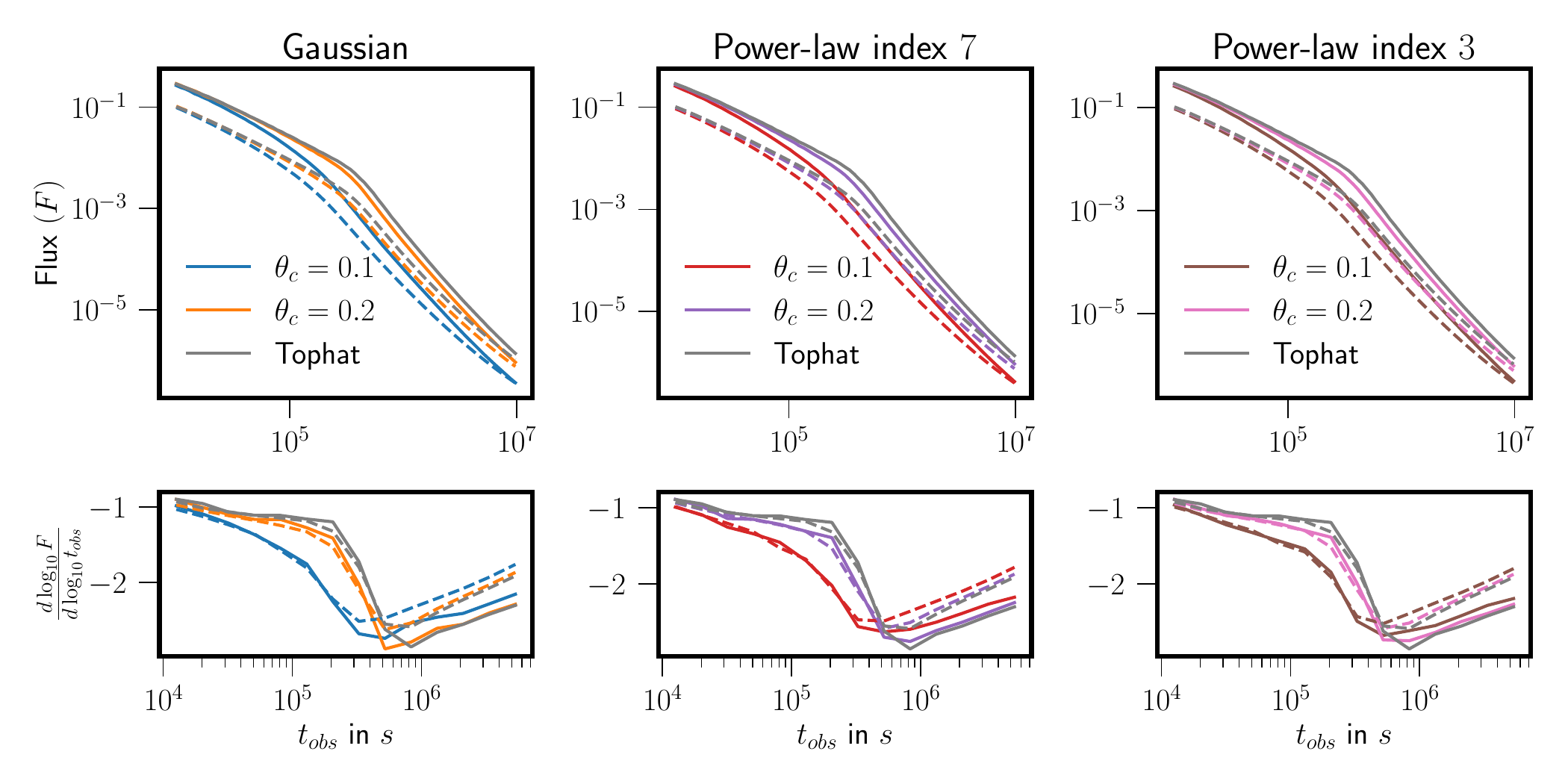}
    \caption{\emph{Top:} On-axis X-ray light curves for different jet structures at a frequency of $10^{17}$\,Hz. The solid lines show results produced using local cooling, while the dashed lines show results using the global cooling approach.
    \emph{Bottom:} The slopes of the light curves shown in the top panel, using corresponding line styles and colours to the top panel.}
    \label{fig:on_axis_lc}
\end{figure*}

\begin{figure*}
    \centering
    \includegraphics[width=\textwidth]{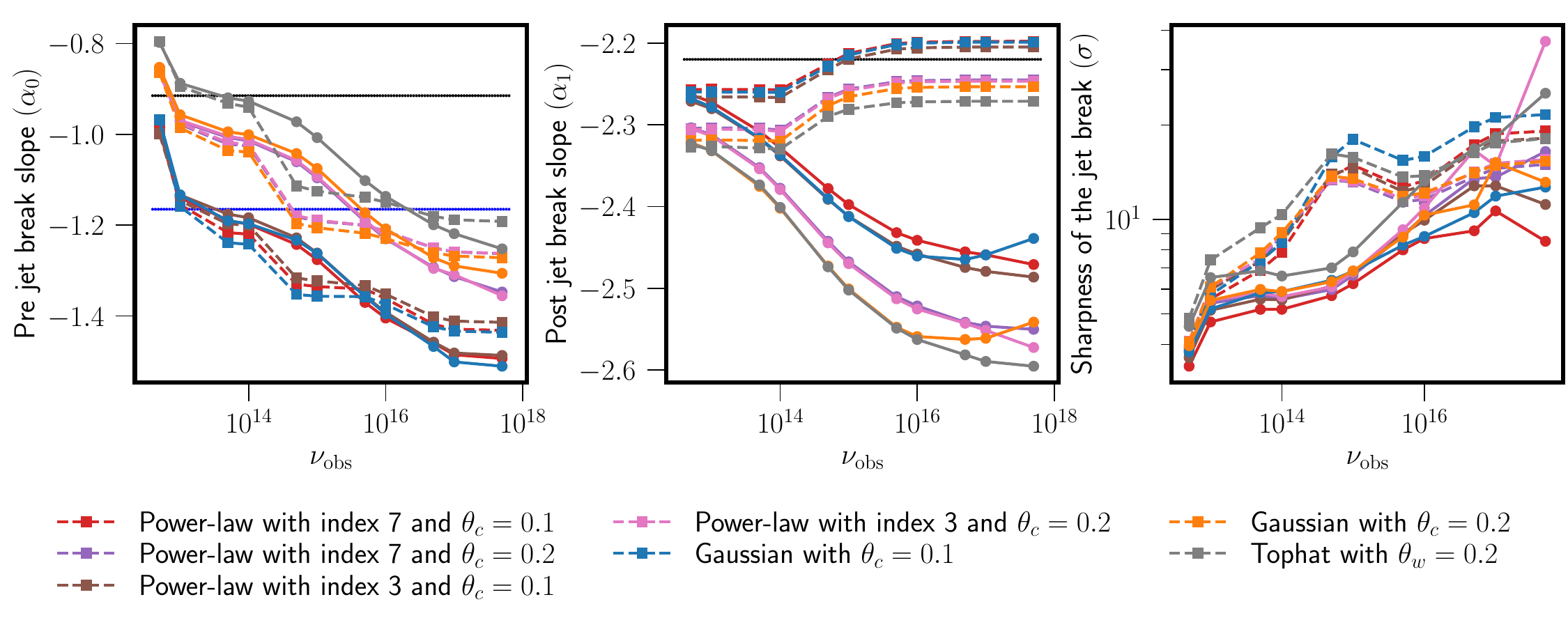}
    \caption{Frequency dependence of the pre-, post-jet break slope and the sharpness of the jet break for different jet structures, considering both cooling scenarios, is shown in the left, middle and right plots, respectively. 
    Solid curves with filled circles are obtained using a local cooling approach, whereas dashed curves with square data points are obtained using a global cooling approach. 
    In the left plot, the analytical prediction of the slopes is shown with black and blue horizontal lines for the cases when $\nu<\nu_{\rm c};\, \alpha_{\rm 0}=3(1-p)/4$  and $\nu>\nu_{\rm c};\,\alpha_{\rm 0}=(2-3p)/4$, respectively. 
    In the middle plot, the analytically predicted slope $(-p)$ is shown with a black horizontal line.}
    \label{fig:break_occur}
\end{figure*} 

We now proceed with an analysis of the afterglow light curves for different jet structures expected for on-axis observations. It has long been known (e.g. \citealt{granot_2001_grba,Kumar_2003, ZhangMacFadyen_2009, vanEerten_2010a}) that the post-break slope deviates from estimates based on simplified modelling of the blast wave dynamics \citep{Rhoads_1997, Sari_1999, Rhoads_1999}. Furthermore, the shape of the transition itself will be sensitive to the structure of the jet, and this has for example been used in recent work to attempt to constrain jet structure from afterglow observations that include a jet break \citep{Lamb2021MNRAS}. It is therefore of interest to take stock of what the impact of the different approaches to electron cooling is on the jet break and post-break light curve slope, if any.

In the top panel of Fig.~\ref{fig:on_axis_lc}, we present the on-axis X-ray lightcurves for different jet structures, as specified in Table~\ref{tab:1}, under two cooling scenarios: global cooling (dashed curves) and local cooling (solid curves). 
For comparison, we also show the light curve of a top-hat jet (grey curves). 
The bottom panel illustrates the slope of these light curves.

All light curves are observed above the cooling break. However, due to the upward shift of the cooling break in a local cooling approach relative to a global cooling approach (as discussed in section \ref{sec:spectral_features}), the flux levels for the former are systematically higher. As can be seen from the slope panels, the post-break slope stays steeper for longer in the local cooling case than in the global cooling case. The time of the on-axis jet break does not show a strong dependency on the jet structure.

We can quantify these statements about the behaviour of the light curve slope across the jet break by fitting smooth transition curve to the numerical results. As with the fits to spectra, we use a smoothly broken power law also used for similar purposes in the past \citep{Beuermann_1999, eerten_2013_boosted}:
\begin{equation}
    F(t_{\rm obs}) = \bar{F}\left[\left(\frac{t_{\rm obs}}{t_{\rm break}}\right)^{-\alpha_{\rm 0}\sigma}+\left(\frac{t_{\rm obs}}{t_{\rm break}}\right)^{-\alpha_{\rm 1}\sigma}\right]^{-1/\sigma},
\end{equation}
where $\alpha_{\rm 0}$ and $\alpha_{\rm 1}$ are the pre-and post-jet break temporal slope respectively and $\sigma$ characterizes the sharpness of the transition.
A higher value of $\sigma$ corresponds to a sharper break, a smaller value reflects a smoother break. The sharpness of the jet break has been argued to provide observable information about the structure of the jet \citep{Lamb2021MNRAS}; it is therefore of interest to verify the extent to which the approach to electron cooling has an impact on what can be inferred from such measurements.

Fig.~\ref{fig:break_occur} shows how $\alpha_{\rm 0}$, $\alpha_{\rm 1}$ and $\sigma$ depend on frequency for different jet structures and for both cooling scenarios. To obtain these results, we have fitted the smooth power law to the numerical results in the time range $3 \times 10^4$ s to $1 \times 10^7$ s, sampling around 90 simulated light curve points, and using a least-squares fit method. All jet break times $t_{\rm break}$ were found lie within the time interval $1 \times 10^5$ to $3.2 \times 10^5$ s.

The leftmost panel shows the pre-break slope $\alpha_{\rm 0}$ according to the fit results. At frequencies below the cooling break, the expected slope is $3(1-p)/4$ and above the cooling break the exepcted slope is $(2-3p)/4$. These results do not show a strong on-axis difference in outcome between local and global cooling approach. The more narrow structured jets struggle to reach the expected asymptotic value within the fitted time interval, remaining steeper than expected.

In the middle panel, we observe the post-jet break temporal slope to demonstrate bifurcation in distinct branches for different cooling scenarios, for $\nu>\nu_{\rm c}$. 
We observe that the slope for all the jet structures deviates from the predicted $-p$ slope, which is shown by the black horizontal line.
For local cooling, the slope becomes steeper with the increase in frequency (solid lines), whereas for global cooling (dashed lines), the slope flattens with the increase in frequency.
Although such a deviation from the predicted slope has already been known for top-hat jets \citep[see][for example]{eerten_2013_boosted}, the distinct branching of the slopes for local and global cooling beyond the cooling break is a notable feature. The split is also not unexpected, because emission above this break is produced by a smaller region in the local cooling approach than in the global approach, which would render the emission more sensitive to changes in jet dynamics.

\begin{figure}
    \centering
    \includegraphics[width=\columnwidth]{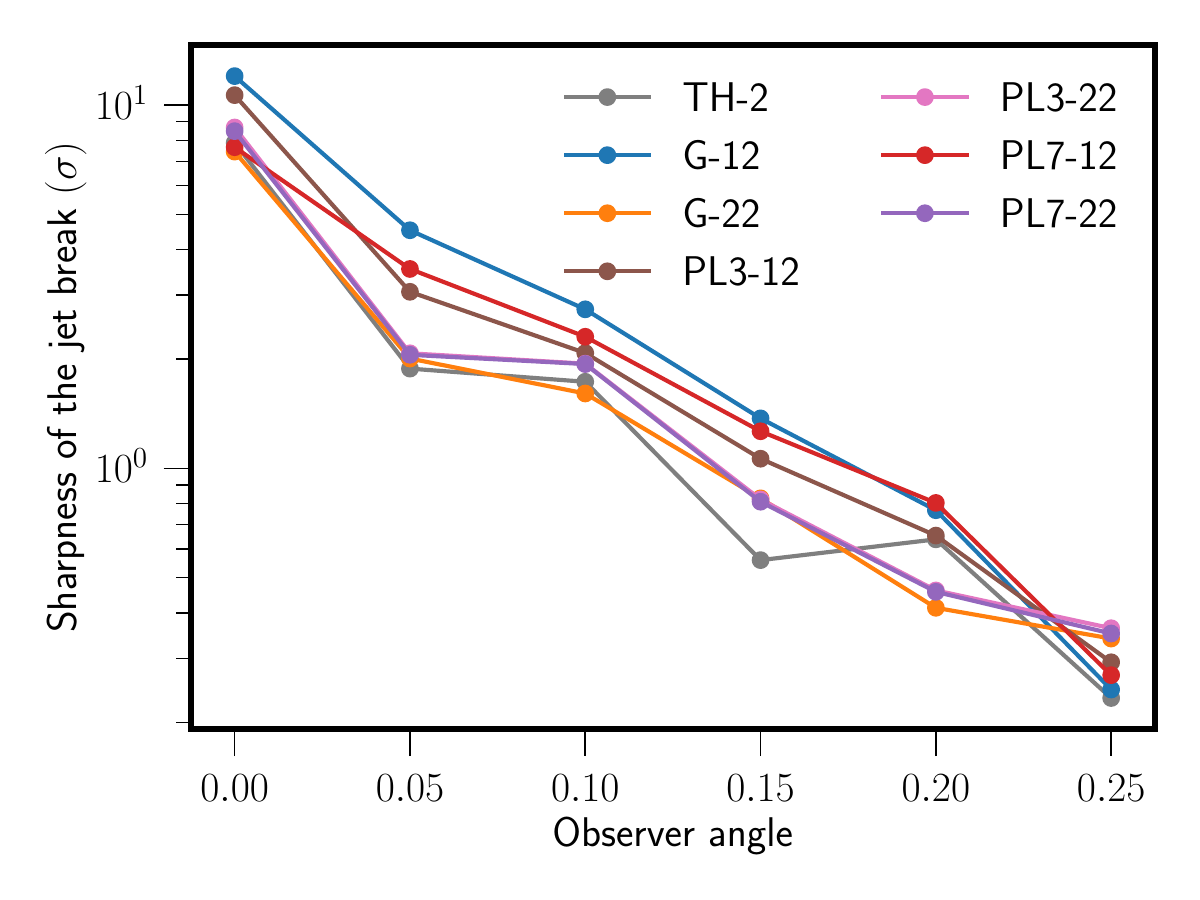}
    \caption{Dependence of the sharpness of the jet-break on the viewing angle at frequency $10^{17}$\,Hz., and for local cooling scenario.}
    \label{fig:off_axis_sharpness}
\end{figure}

In the rightmost panel, the temporal sharpness parameter is plotted logarithmically for different jet structures.
The jet break generally becomes sharper with increasing frequency, again as expected when limb-brightening becomes more pronounced. On the other hand, because the cooling break is shifted upwards in a local cooling approach relative to its global cooling value, this trend is likewise as a whole shifted and there is some indication that the sharpness of the cooling break becomes \textit{less} when comparing identical frequencies between local and global cooling. This difference is at least of comparable order to the impact of jet structure on the sharpness of the transition. 
A further complicating factor is the dependency of the sharpness on observer angle, as illustrated for an observer frequency of $10^{17}$ Hz and various jet structures in Fig. \ref{fig:off_axis_sharpness} and noted e.g. by \cite{granot_2001_grba,DeColle2012ApJdynamics,Lamb2021MNRAS}.
Caution should therefore be applied when interpreting observations of jet break transitions directly in terms of jet structure.

\section{Discussion}\label{sec:discuss}

In this work we have focused on a set of standardized templates for lateral jet structure (Gaussian, power-law). There is compelling observational support for a Gaussian-like structure for at least GRB 170817A (e.g. \citealt{Troja_2017, Margutti2017ApJ, Hallinan_2017, Mooley_2018, LambKobayashi2018MNRAS,gill_2018, Ghirlanda2019, Troja_2019, Troja_2020, Ryan_2020}), whereas for example GRB 221009A has been modelled using a power-law structure \citep{o'Connor_2023}. These particular functional forms though are mere templates of some practical utility to capture an intrinsic structure that can well be more rich (see \citealt{TakahashiIoka2020MNRAS} for an analysis that sidesteps the use of pre-constrained templates), and the results presented in this work should be similarly taken as such.

Physically, the structure of GRB jets is shaped by a number of factors, ranging from the structure imposed during launching of the jet and interaction with the accretion disk (e.g. \citealt{Kathirgamaraju2019}) to the subsequent interaction with any confining medium (e.g. a stellar envelope in the case of a collapsar origin, \citealt{Mizuta2013}, or merger debris or neutrino-driven wind for a neutron star merger, \citealt{Geng2019}). 
The final structure of the GRB jets also depends on the turbulent mixing of material between the jet and cocoon.
Detailed 3D simulations show that turbulent jet–cocoon mixing is intrinsically asymmetric, yet for afterglow modelling the jet can be reasonably approximated with an axi-symmetric angular distribution \citep{gottlieb_2021}. 
As a physical underpinning of this approximation, after breakout, the cocoon pressure falls sharply and the jet self-organises into a lateral energy profile \citep{Lazzati2005}. This implies that our imposition of axi-symmetry in the simulations is not unrealistic.

For shock acceleration we have adopted an instantaneous acceleration model, resulting in a power-law like non-thermal particle distribution. It is an approximation to assume this power-law remains intact between $\gamma_{\rm min}$ and $\gamma_{\rm max}$; more realistically there would be no instantaneous drop at $\gamma_{\rm \max}$, but rather a steep turnover (as e.g. demonstrated in \citealt{GranotSari_2002}). Adopting a more accurate description here however would only have minor impact on our conclusions, including those about the sharpness of the cooling break in the spectrum and the temporal smoothness of the jet break, given the rapidness of the turnover as $\gamma_{\rm max}$ is approached. Similarly, the difference between incorporating a smooth synchrotron spectrum \citep{GranotSari_2002} and a series of connected power laws (this work), will lead to a smoother cooling-break transitions for the former, leaving the more stark discrepancey between local and global cooling intact.

When assessing the sharpness of the spectra, the temporal evolution of the characteristic spectral features (in particular the cooling break $\nu_{\rm c}$) and the temporal transition of light curves across the jet breaks, we have made use of ad-hoc fits to our simulation output with smooth power law fit functions. Inevitably, the results of such fits will be shaped in part by aspects other than the underlying model, such as the numerical noise of the simulation, the  number of simulation data points included in the fit and the spectral / temporal range used for the fit. Furthermore, any conclusion based on a fit function will be subject to whatever limitation is inherent in the fit function itself. In our case, the number of data points used for our fits (spectral and temporal feature fits in Section \ref{sec:spectral_features}; temporal feature fits in Section \ref{sec:light_curves}) is large enough not to skew the fit results, while the smooth power-law fit functions that we have used should allow for at least a direct comparison between the current work and the literature where these were previously applied. As stated in Section \ref{sec:spectral_features}, the presence of numerical noise implies an error bar on our fit results. The lack of a long period of true asymptotic behaviour shown in the evolution of the characteristic features of the spectrum is as expected from simulations but will introduce a skew in any slope assessment based on a fit function. On the other hand, actual data from observations will present highly analogous challenges to model fitting, including e.g. a lack of true asymptotic behaviour, and skew the output of a fit in a similar manner.

Finally, in this study we do not take into account additional cooling due to inverse Compton scattering, which lies beyond the scope of our numerical approach. Inverse Compton scattering would shift the cooling break towards lower frequency to an extent largely independent of our approach (local or global) to synchrotron cooling. We therefore expect our conclusions about the relative differences between the two to carry over if inverse Compton scattering were included. We do note that our results suggest caution when interpreting an observed cooling break: it is the difference between observation and expectation from the more accurate \emph{local} cooling approach that matters for the purpose of establishing how much additional cooling is inferred to have been caused by inverse Compton scattering .

\section{Summary and conclusions}\label{sec:conclusion}

In this work, we have studied the dynamical evolution of relativistic structured jets decelerating in a constant density environment. We have set up jets using Gaussian, power-law and top-hat distributions of outflow energy with angle. Additionally, we have locally traced the shock-accelerated electron populations and examined the impact of two different methods to account for synchrotron cooling: a ``global'' approach where a single cooling time (an adiabatic expansion time, taken to be the time since explosion) is used to estimate the shape of the electron spectrum and a ``local'' cooling approach where the upper cut-off electron Lorentz factor is traced explicitly over time through an advection equation. The simulations were performed in two-dimensional axi-symmetry using our moving-mesh code \textsc{gamma}, an approach that made it possible to obtain the high spatial resolution necessary to track the rapidly evolving electron spectrum to sufficient accuracy. Altogether, we arrive at the following conclusions:
\begin{itemize}
\item The  temporal evolution of the Lorentz factor at the tip of the jet is influenced by the lateral structure of the outflow, leading to a smoother and more extended transition across the dynamical jet break for structured jets than for top-hat jets (Fig. \ref{fig:tip_Lorentz_factor}). During this process, the tip Lorentz factor will drop more steeply for structured jets than for top-hat jets. This demonstrates the additional impact of the initial lateral pressure gradient in structured jets, on top of the increase in jet spreading triggered by the 
lateral rarefaction wave from the jet edge reaching the jet axis for both structured and top-hat jets.
\item Tracing the lateral motion of a sound wave along the jet surface remains a good predictor for the full turnover to jet spreading even for structured jets, and key to understanding why a regime of relativistic lateral spreading is rarely achieved in practice. As with the jet tip behaviour, structured jets experience a more extended transition towards lateral spreading (Compare Figs. \ref{fig:theta_track_Gaussian} and \ref{fig:theta_track_top}). 
\item In a local approach to electron cooling, the resulting cooling break in the spectrum is shifted to significantly higher frequencies relative to a global approach (see also \citealt{Ayache_2022}). This difference in $\nu_{\rm c}$ between local and global cooling is robust across jet structures and more extreme than any additional shift induced by jet structure (compare Figs. \ref{fig:cool_evol_loc} and \ref{fig:cool_evol_glob}).
\item The spectral transition around the cooling break will be smoother for a local approach to electron cooling than for a global approach  (see Fig. \ref{fig:cool_evol_sharp}). All structured jets have near-identical cooling break sharpness ($s_{\rm c} \sim 4$ for global cooling, $s_{\rm c} \sim 0.9$ for local cooling).
\item For local cooling, the spatial extent of the region behind the shock front dominating the emission emission region will be highly sensitive to observer frequency, for frequencies above the cooling break. This feature persists across time, as expected. Different jet structures produce differently-sized emission regions, but the impact of structure is minor relative to the difference between global and local cooling (Fig. \ref{fig:rbar_loc}).
\item Due to limb-brightening at high observer frequencies being more pronounced for local cooling, the post-break temporal slope at such frequencies (e.g. X-rays) will at least initially be steeper in a local cooling approach than in a global cooling approach (Figs. \ref{fig:on_axis_lc} and \ref{fig:break_occur}). 
\item Both for local and global cooling, the jet break generally becomes sharper with increasing frequency relative to the cooling break. The impact of global versus local cooling is at least of comparable order to that of jet structure, suggesting caution when interpreting observed jet breaks in terms of jet structure through a model that does not fully account for the local nature of electron cooling (see Fig. \ref{fig:break_occur}).  
\end{itemize}

\section*{Acknowledgements}

The authors thank the anonymous referee for the helpful comments and constructive remarks on this manuscript. The authors gratefully acknowledge support by the Science and Technology Facilities Council (STFC) through grant ST/X001067/1. HJvE further acknowledges support by the European Union Horizon 2020 programme under the AHEAD2020 project (grant agreement number 871158). This work used the Isambard 2 UK National Tier-2 HPC Service (http://gw4.ac.uk/isambard/) operated by GW4 and the UK Met Office, and funded by EPSRC (EP/T022078/1).

%%%%%%%%%%%%%%%%%%%%%%%%%%%%%%%%%%%%%%%%%%%%%%%%%%
\section*{Data Availability}

The derived data generated in this research will be shared on reasonable request to the corresponding author. All data has been generated using the open-source code \textsc{gamma}.



\bibliographystyle{mnras}
\bibliography{paper} 
%%%%%%%%%%%%%%%%%%%%%%%%%%%%%%%%%%%%%%%%%%%%%%%%%%

%%%%%%%%%%%%%% APPENDICES %%%%%%%%%%%%

\appendix

\section{Further analysis of long-term evolution of spherical and wide-angle jets}
\label{appendix:wide_jets}

\begin{figure}
    \centering
    \includegraphics[width=\columnwidth]{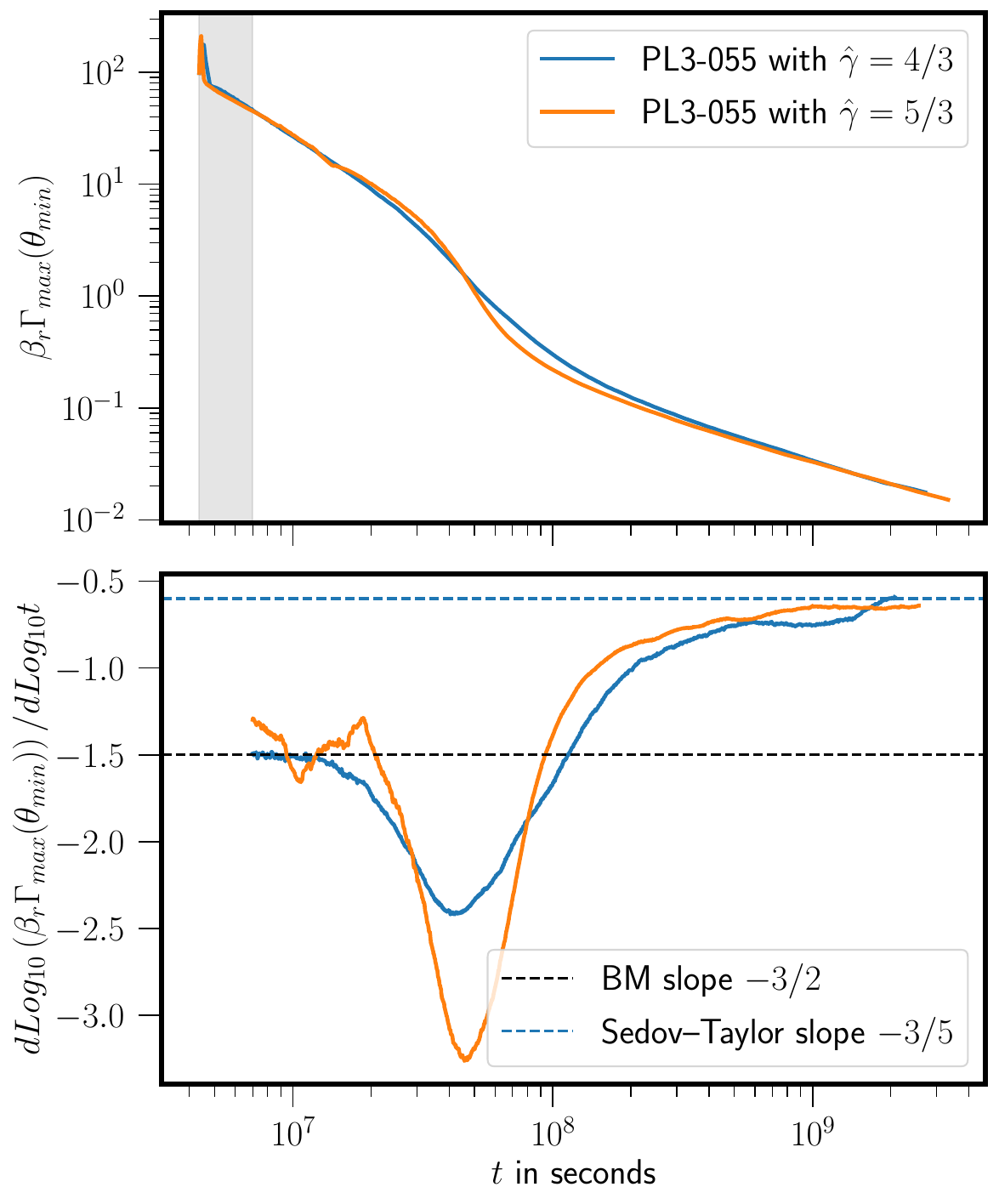}
    \caption{Temporal dynamics of the broad-wing power-law jet PL3-055, converging onto the Sedov-Taylor regime.}
    \label{fig:ST}
\end{figure}

We use our wide-angle run PL3-055 to further investigate the long-term evolution of structured jets and the convergence onto the late-time Sedov-Taylor regime, the role of $\theta_{\rm w} / \theta_{\rm c}$ and our choice of adiabatic exponent.

Figure \ref{fig:ST} shows the eventual convergence of the jet onto the Sedov-Taylor regime. We include two cases, differing in their value for adiabatic exponent $\hat{\gamma}$. While the velocity at the tip of the jet reaches the non-relativistic asymptote for dynamical evolution at comparable time for both $\hat{\gamma} = 4/3$ and $\hat{\gamma} = 5/3$, it is interesting to note that the non-relativistic value leads to a smoother transition.

\begin{figure}
    \centering
    \includegraphics[width=\columnwidth]{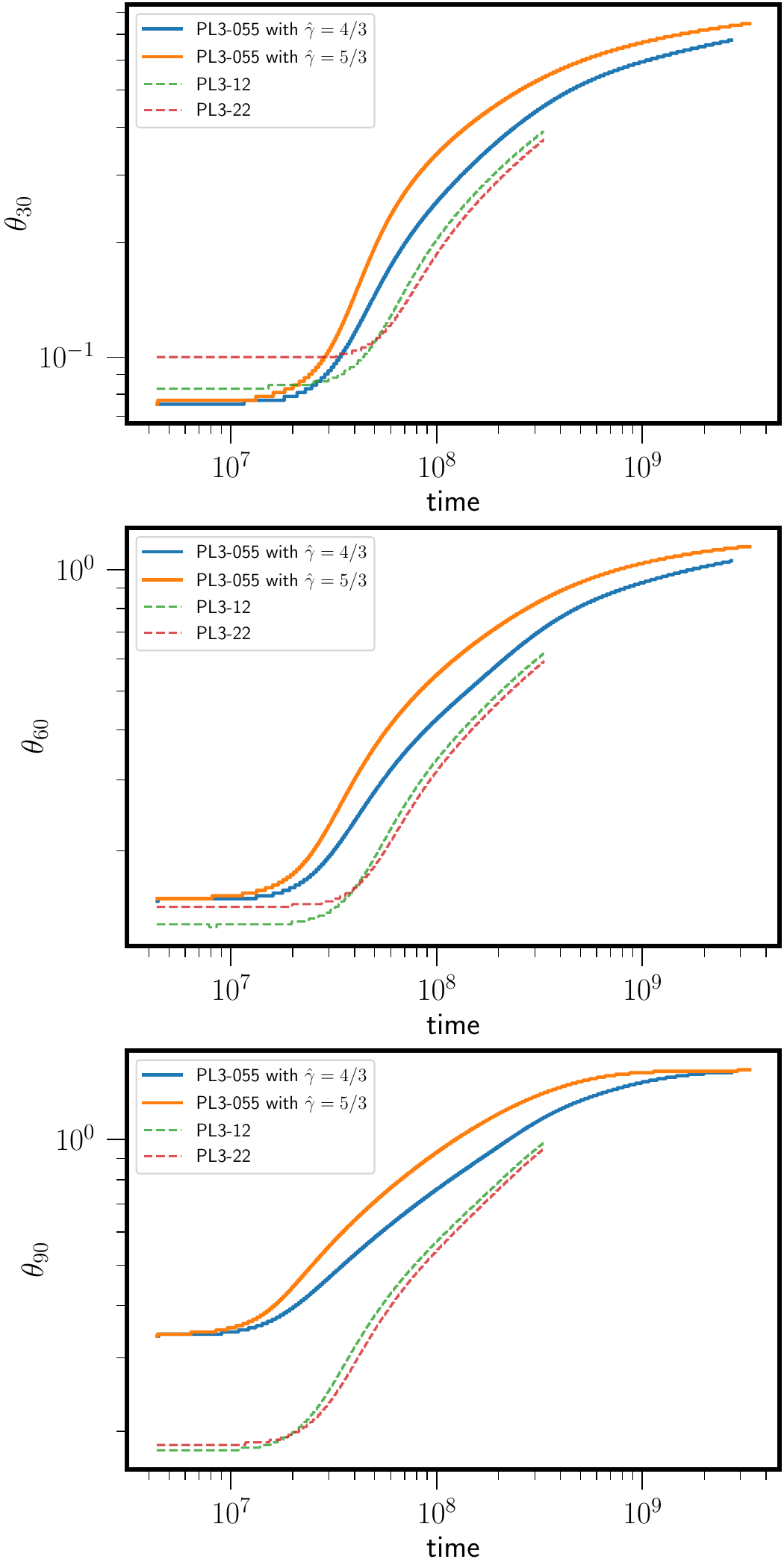}
    \caption{Angular spreading diagnostics broader-wing power-law jet ($n=3$, $\theta_c=0.05$, $\theta_w=0.5$) and comparison with the paper-run of same jet structure but with $\theta_c=0.1\, \&\, 0.2$, $\theta_w=0.2$. See Table \ref{tab:1} for a description of the jet set-ups.}
    \label{fig:spread}
\end{figure}

This difference is also seen in Fig. \ref{fig:spread}, showing the spreading of the jet to start earlier for $\hat{\gamma} = 5/3$ but then following the same intermediate asymptote as the $\hat{\gamma} = 4/3$ case. The fixed adiabatic exponent simulations will bracket the spreading dynamics of a jet with an EOS that incorporates a changes adiabatic exponent with time.

\begin{figure}
    \centering
    \includegraphics[width=\linewidth]{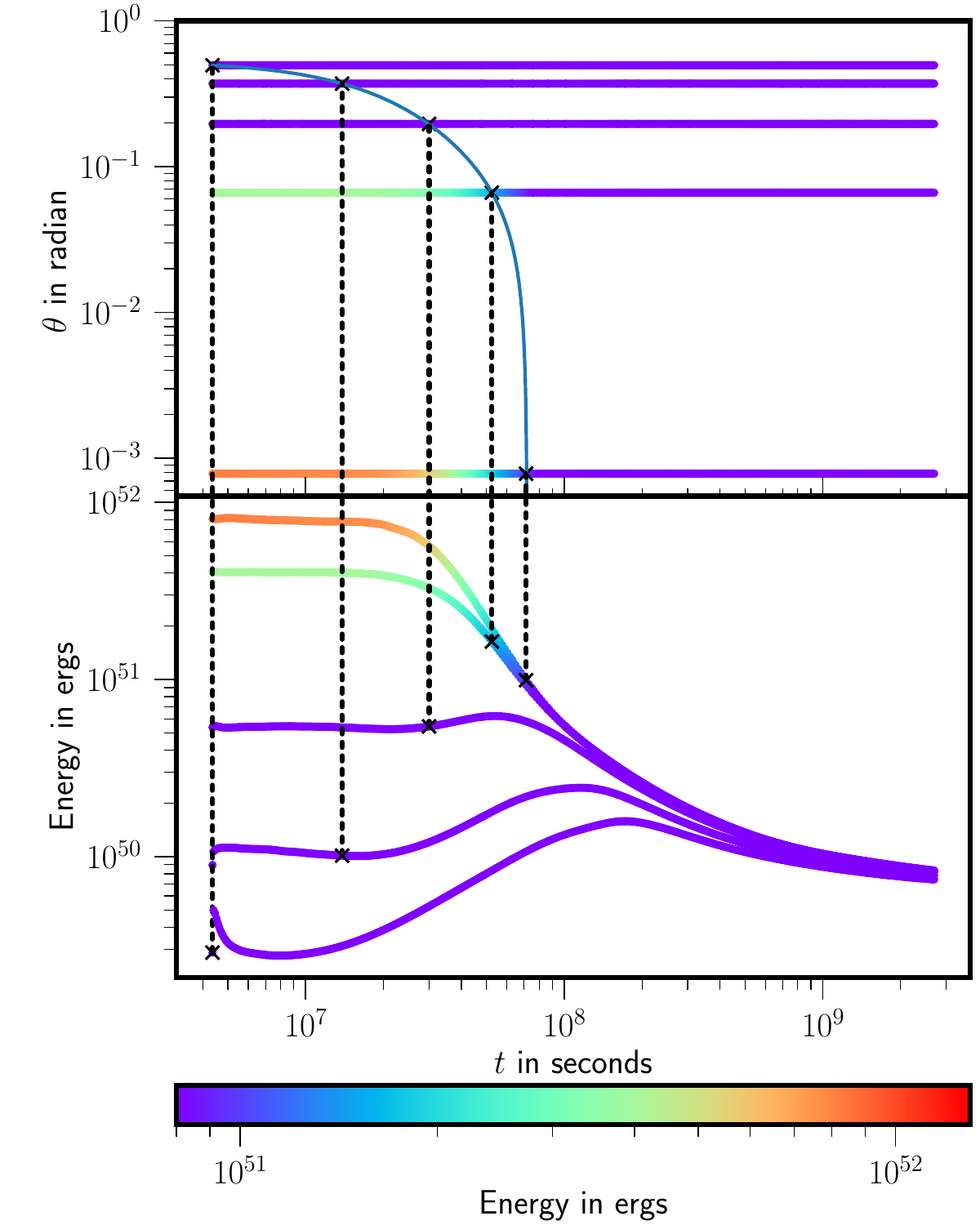}
    \caption{Evolution of the lateral distribution of energy for the broad-wing power-law jet PL3-055, with $\hat{\gamma} = 4/3$, compared to a sound wave traveling across the surface of the jet indicating the inward motion of a rarefaction wave.}
    \label{fig:rarefaction}
\end{figure}

In Fig. \ref{fig:rarefaction} we show the redistribution of energy over time for our broad-wing simulation. The same comparison to a sound wave running along the shock front as demonstrated in Figs. \ref{fig:theta_track_Gaussian} and \ref{fig:theta_track_top} is shown. The sound wave position estimate based on relativistic flow and accounting only for the fluid immediately downstream, can be seen to overestimate the onset of the turning point for angles near the tip (shown most clearly in the bottom panel of the figure). Of course, broad jets would not be expected to exhibit a regime of relativistic spreading to begin with, regardless of jet structure.

\section{Lateral sound waves}
\label{appendix:sound_waves}

In the ultra-relativistic limit, the squared magnitude of the speed of sound in the local fluid frame is given by
\begin{equation}
\left(\beta'_{\rm s}\right)^2 = \frac{1}{3} = \left( \beta'_{\rm s, \uptheta} \right)^2 + \left( \beta'_{\rm s, r}\right)^2,
\label{eq:sound_wave}
\end{equation} when expressed as the sum of squared components in the radial and angular direction. If this sound wave is to progress along the shock front, the sound wave velocity radial component in the lab frame needs to remain equal to the shock velocity in the lab frame $\beta_{\rm SH}$, i.e. $\beta_{\rm s,r} = \beta_{\rm SH}$. For a fluid moving radially with Lorentz factor $\gamma$ (and light speed-normalized velocity $\beta$), the following transformation rules are relevant:
\begin{equation}
\beta'_{\rm s,r} = \frac{\beta_{\rm s,r} - \beta}{1 - \beta \beta_{\rm s,r}}, \qquad \beta_{\rm s, \uptheta}' = \frac{\beta_{\rm s, \uptheta}}{\gamma \left( 1 - \beta \beta_{\rm s, r}\right)}
\label{eq:velocity_transformation}
\end{equation}
According to Equations \ref{eq:BM}, we have the relation $\Gamma_{\rm S}^2 = 2 \gamma^2$ between the shock Lorentz factor $\Gamma_{\rm S}$ and the fluid Lorentz factor directly behind the shock front. If we use this in Equation \ref{eq:sound_wave}, along with the velocity transformation rules from Equations \ref{eq:velocity_transformation}, and solve for $\beta_{\rm s, \uptheta}$, we find
\begin{equation}
\beta_{\rm s, \uptheta} = -\frac{1}{2 \Gamma_{\rm S}}.
\end{equation}
Therefore, the evolution of a sound wave from the edge to the tip of the jet is given by
\begin{equation}
R\frac{d \theta}{dt} = c \beta_{\rm s, \uptheta} \Rightarrow \frac{d \theta}{d t} \approx -\frac{1}{2 \Gamma_{\rm S} t}.
\end{equation}
At this point we have used the assumption of ultra-relativistic flow, but have not yet included anything about jet structure or time evolution of $\Gamma_{\rm S}$. In the case of a top-hat jet of width $\theta_{\rm w}$ evolving according to $\Gamma^2 \propto t^{3-k}$ (i.e. according to the BM solution in an environment with external density depending on radius as $r^{-k}$) and a simulation picking up the point explosion solution at $\Gamma_{\rm S} = \Gamma_{\textrm{S},0}$, we get
\begin{equation}
\theta_{\rm w} - \theta = \frac{1}{\left( 3 - k\right) \Gamma_{\rm S}} - \frac{1}{\left( 3 - k \right) \Gamma_{\textrm{S},0}}.
\end{equation}
The limiting case with $\Gamma_{\textrm{S},0} \to \infty$ would then have its lateral rarefaction wave reaching the tip at $\Gamma_{\rm S} = \theta_{\rm w}^{-1} / \left( 3 - k\right)$, or $\gamma \approx 0.24 \theta_{\rm w}^{-1}$ for the ISM case. The numerical pre-factor $0.24$ helps to explain why `typical' jets with $\theta_{\rm w} \sim 0.1$ rad do not go through an asymptotic regime of ultra-relativistic spreading, which would have required $\gamma \gg 2.4$.

We note that our sound wave prescription based on the BM solution is inherently assuming the radial motion of the shock front to be relativistic. This is a reasonable assumption as long as $\theta_{\rm w}$ is small enough for the sound wave to reach the tip while $\Gamma_{\rm tip} \gtrsim 2$, as is the case in the simulations emphasized in this work. It is also exactly on target for predicting the turn-over in Fig. \ref{fig:tip_Lorentz_factor}. However, the assumption of relativistic flow can be relaxed in the same manner that jet spreading models have included trans-relativistic flow regimes, as done e.g. in \cite{GranotKumar2003} (and which reduces to a near-identical result in the relativistic limit, even if the difference between shock front and shocked plasma radial motion is not taken into account in their one-zone approach). In the non-relativistic limit the angular position of a sound wave changes in a logarithmic manner \citep{vanEerten_2010a}. However, ultimately, as discussed in Section \ref{section:jet_spreading}, the drop in energy content along a given outflow angle over time is affected more by the full radial profile of the flow than by the shock front segueing into the trans-relativistic regime, and made more smooth as a result.

\section{Global cooling and the edge of the jet}
\label{appendix:shearing_layer}

As shown in Figure \ref{fig:cool_evol_sharp}, for the global cooling scenario, there is an offset between the sharpness of the structured jets and spherical blast wave. Since this is visible already before the nominal ``jet break'' in the light curve including for top-hat jets, it deserves further commenting on, even if an artifact of a global approximation to electron cooling. The offset can be attributed to a shearing layer that is present throughout the multi-dimensional evolution of any jet with finite opening angle. 
Within this layer, the Lorentz factor, which defines the beaming of the radiation, can become lower than the tip Lorentz factor, putting these regions within the field of view of the observer already early on.

\begin{figure}
    \centering
    \includegraphics[width=\columnwidth]{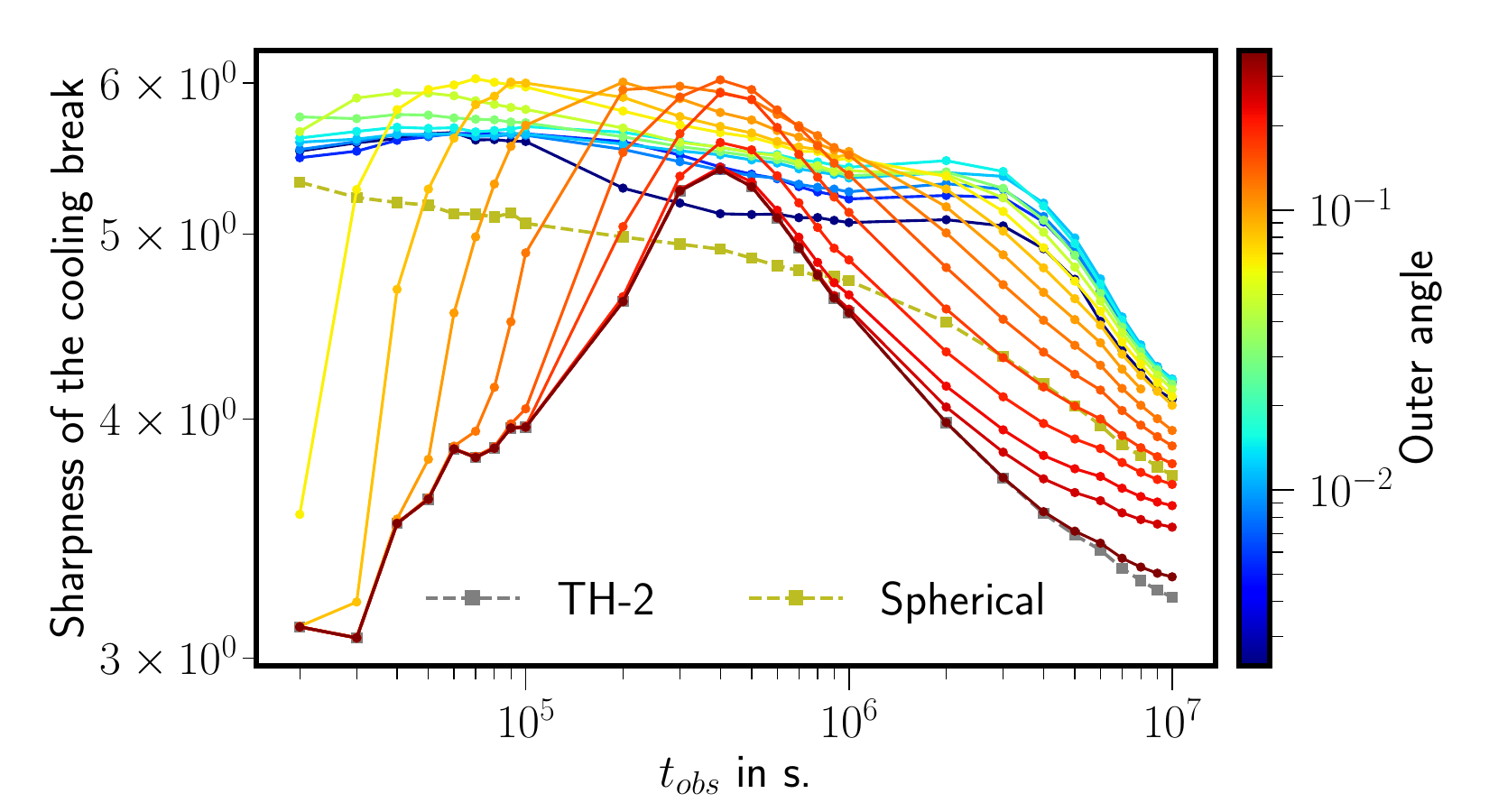}
    \caption{Dependence of the evolution of the sharpness of the cooling break on how much of a top-hat jet (TH-2) is included in the calculation. A global cooling scenario is assumed and the different colours represent computations based on different cut-off points in emission angle when computing the flux (i.e. dark blue is a narrow conic edge entirely within the top-hat jet, dark red a full sphere extending beyond the top-hat).
    % \hjve{The ``TopHat'' label needs improving or removing altogether} 
    }
    \label{fig:top_sharp}
\end{figure}

To demonstrate this behaviour, in Fig~\ref{fig:top_sharp} we show the evolution of the sharpness of the cooling break from the cumulative emission up to a truncation angle $\theta$ measured from the tip of the jet. 
The figure clearly shows that if larger angles are excluded from the emission for our top-hat TH-2 model, the sharpness evolution does not yet show its peaked structure and resembles more closely the spherical version. This is as expected, since the inner parts of the jetted outflow are still expected to more radially outwards, like their spherical model counterparts.
However, when progressively higher angles are included in the emission, we can see the sharpness falls down at earlier and starts to converge to the top-hat jet result. The change becomes pronounced when the exclusion angle begins to approach the truncation angle of the actual jet, pointing at the impact of the material in the shearing layer found in this angular region.

%%%%%%%%%%%%%%%%%%%%%%%%%%%%%%%%%%%%%%%%%%%%%%%%%%

% Don't change these lines
\bsp	% typesetting comment
\label{lastpage}
\end{document}